 \documentclass[reprint]{JASA}

\usepackage[mathscr]{eucal}
\usepackage{mathtools}

\newtheorem{remark}{Remark}

\newcommand{\kk}{\mathbf k}
\newcommand{\ffoot}[1]{}
\definecolor{myviolet}{rgb}{0.6,0.2,1}
\newcommand{\fff}{{\mathscr F}} 
\newcommand{\equalexpl}[1]{
  \underset{\substack{\uparrow\\\mathrlap{\text{\hspace{-1em}#1}}}}{=}}

\usepackage{xcolor}
\usepackage[normalem]{ulem}
\newcommand{\dario}[1]{{\color{green!50!black} #1}}
\usepackage{capt-of}

\begin{document}

\title[The sub-harmonic series via a linear model of the cochlea]{A solution to the mystery of the sub-harmonic series
  via a linear  model of the cochlea
}
\author{Ugo Boscain}
\email{ugo.boscain@sorbonne-universite.fr }
\affiliation{LJLL, Sorbonne Universit\'e, Universit\'e de Paris, CNRS, Inria, Paris, France.}
\author{Xiangyu Ma}
\email{xiangyu.ma@sorbonne-universite.fr}
\affiliation{LJLL, Sorbonne Universit\'e, Universit\'e de Paris, CNRS, Inria, Paris, France.}
\author{Dario Prandi}
\email{dario.prandi@centralesupelec.fr}
\affiliation{L2S, Universite Paris-Saclay, CentraleSup\'elec,  CNRS,
  Gif-sur-Yvette, France.}
\author{Giuseppina Turco}
\email{giuseppina.turco@cnrs.fr}
\affiliation{LLF, Université Paris Cité, CNRS, Paris, France.
\\[3mm]
{\em ~~~~~~~~~~~~~~~~~~~~~~~~~~~~~The authors contributed equally to this work.}
}


\date{\today}

\begin{abstract}
  In this paper, we study a simple linear model of the cochlea as a set of vibrating strings. We make the hypothesis that the information sent to the auditory cortex is the energy stored in the strings and consider all oscillation modes of the strings. We show the emergence of the sub-harmonic series whose existence was hypothesized in the XVI century to explain the consonance of the minor chord. We additionally show how the nonlinearity of the energy can be used to study the emergence of the combination tone (Tartini’s third sound) shedding  new light  on this long-debated subject.
\end{abstract}


\maketitle


\section{Introduction}

One of the most intriguing concepts in music theory is the so-called  {\it sub-harmonic  series}, also referred  to the undertone series  or hypotonic series, which is defined in a way that mirrors the well-known {\it harmonic series}  (also called overtones series).
%

\begin{figure}
  \begin{center}
    \includegraphics[width=5.5truecm]{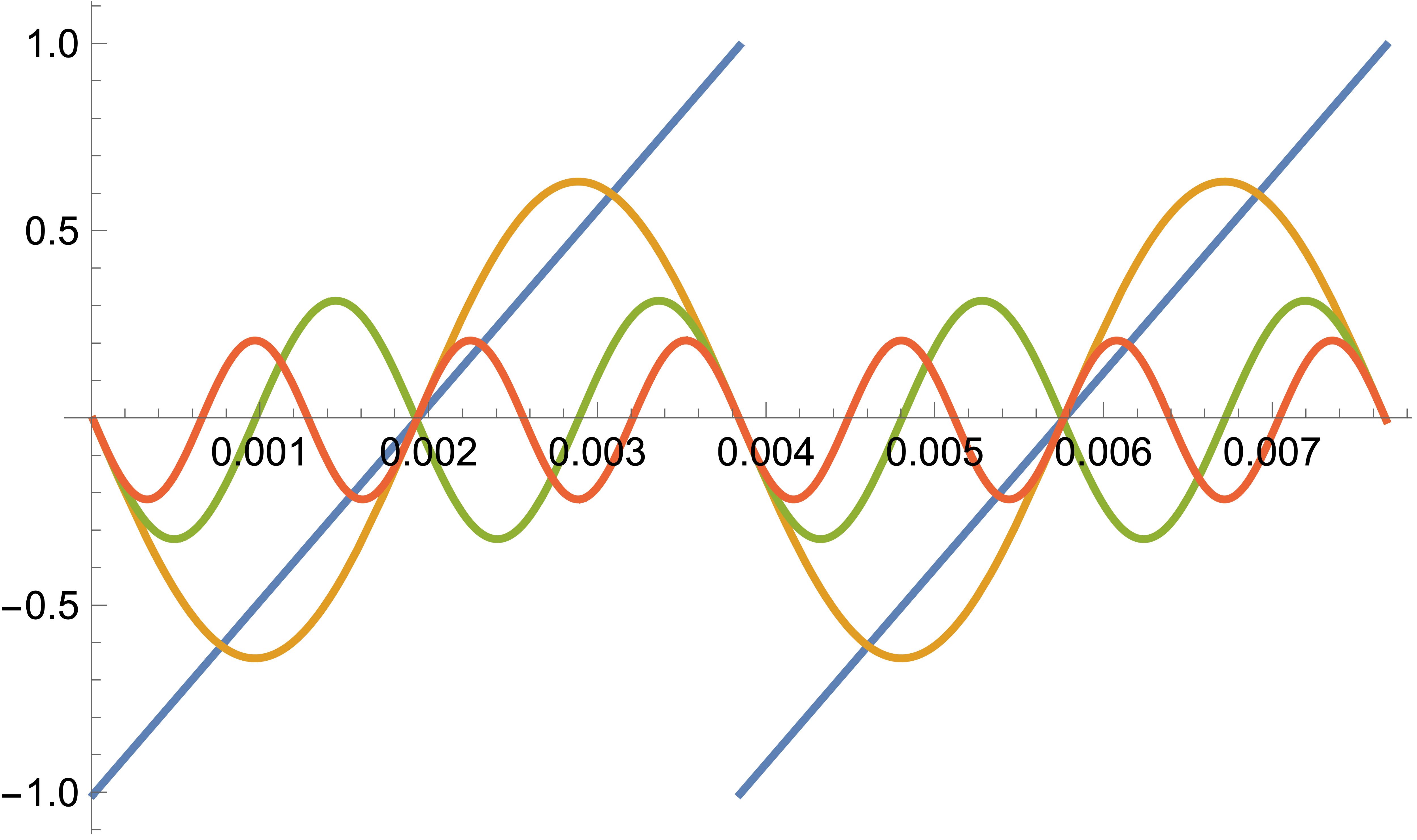}\\[2mm] \includegraphics[width=5.5truecm]{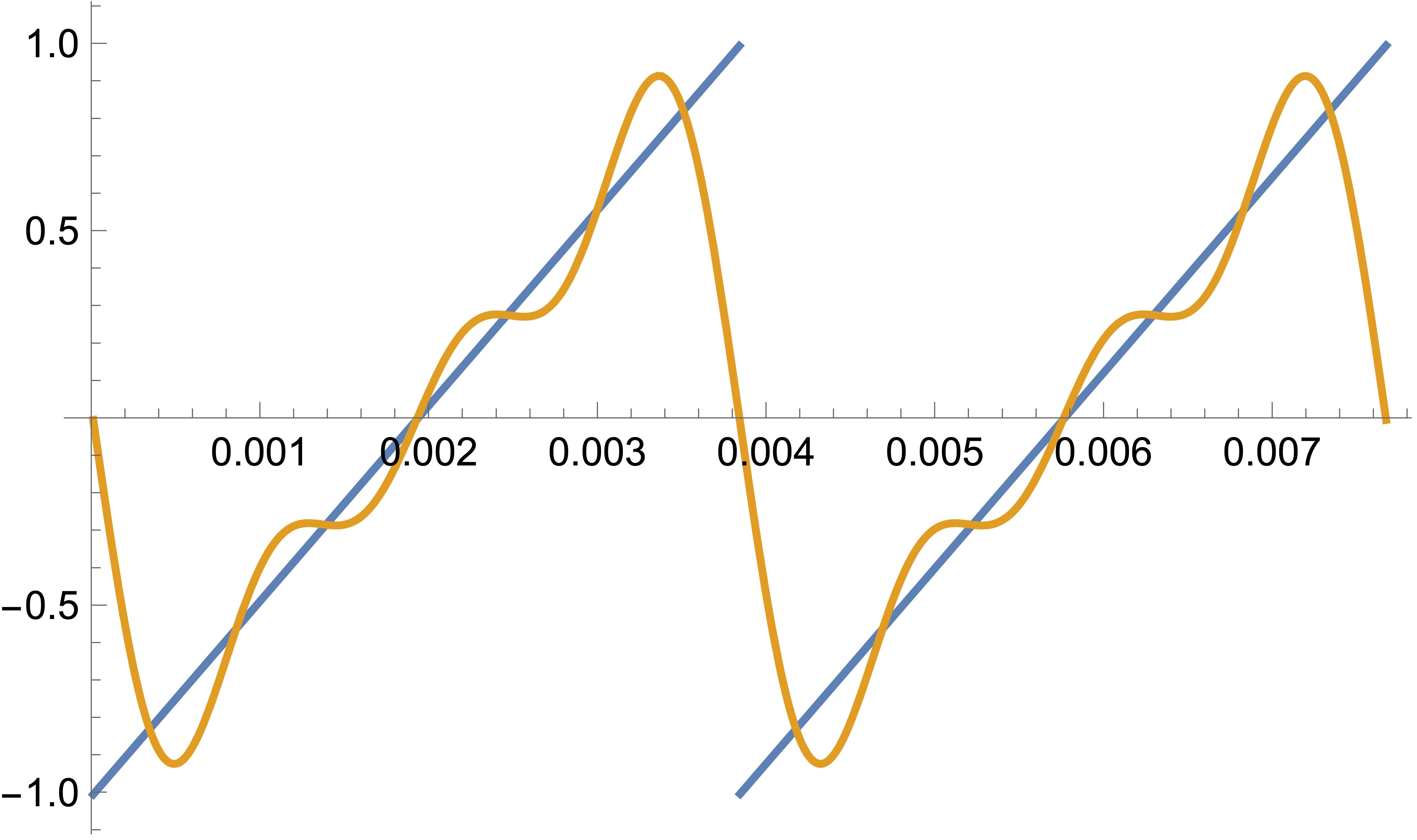}
    \caption{Two periods of a sawtooth signal  with a frequency of 262Hz (in blue) with its first three Fourier components (top) and with their sum (bottom)\label{f-dente-di-sega-Fourier}.}
  \end{center}
\end{figure}

The {\em harmonic series} is well understood from a mathematical, physical, and psychoacoustic perspective. A harmonic sound of frequency $\fff$ (i.e., a \emph{periodic} variation in air pressure of period $\tau=1/\fff$) can be decomposed into sinusoidal components at frequencies that are integer\footnote{Throughout this paper, the set of integers is understood not to include zero.} multiples of $\fff$ (see Figure \ref{f-dente-di-sega-Fourier}), possibly with phase differences. This decomposition is well established physically (e.g., a piano string vibrates not only in its fundamental mode but also in higher modes), mathematically (Fourier series), and perceptually: the presence or absence of the various harmonic components determines the perception of the timbre of the original harmonic sound.

The $n$-th harmonic refers to the Fourier component at frequency $n\fff$.
For instance, if the fundamental frequency of a harmonic sound\footnote{In Scientific Pitch Notation (SPN), see \citep{SPN}.}
corresponds to C$_4$ ($\fff = 262$Hz), the harmonic  series will produce the notes C$_5$ (the second harmonic of frequency $2\fff$), G$_5$ (the third harmonic of frequency $3\fff$), C$_6$ (the fourth harmonic of frequency $4\fff$), E$_6$ (the fifth harmonic of frequency $5\fff$), and so on—these are exactly the notes playable on a piccolo trumpet in C without engaging any valve.\footnote{It is important to note that knowledge of the intensity of the harmonics of a harmonic sound  is not sufficient to reconstruct the original sound. In fact, such reconstruction also requires knowledge of the relative phases. Interestingly, the human auditory system is generally insensitive to these relative phases for harmonic sounds (although they can be perceived for more complex sounds)

  \citep{Moore2012,ZwickerFastl1999}.}

Since the birth of music theory, one of the most challenging problems has been to understand why certain combinations of harmonic sounds are perceived as being consonant or dissonant, and whether this perception is shaped by cultural factors or by underlying physical phenomena. The presence of certain intervals (such as the octave, the fifth, etc.) across many very different cultures has led to the idea that at least some basic intervals are perceived as consonant for physical reasons.
This point of view was adopted by Zarlino (1517-1590) \citep{Zarlino1558}, Descartes (1596-1650) \citep{descartes1650},
Rameau (1683-1764) \citep{rameau1722}, Lagrange
(1736- 1813) \citep{Lagrange1759,Lagrange1762}, Helmholtz (1821-1894) \citep{Helmholtz1863},  (see also \citep{pamcutt1989harmony,Steege2012,Castellengo2015}  for a historical perspective).

The major triad is one of the most relevant chords whose consonance has been justified through this idea. Indeed, all notes forming the triad appear in the harmonic series of the root. For instance, the C major triad (C–E–G) arises from the first five harmonics of a root of C. The hypothesis, therefore, is that the major triad is perceived as consonant because the auditory cortex is accustomed to encountering it within harmonic sounds that are sufficiently rich in overtones.
The question of consonance of the minor triad is more subtle. A possible explanation was already put forward
in \citep{Zarlino1558}. In this treatis, Zarlino hypothesized  the existence of the  {\em sub-harmonic series}: the mirror image of the harmonic series, composed of notes obtained by dividing the fundamental frequency by 2, 3, 4, etc. If we take again C$_4$ ($\fff=262$Hz) as the reference, the  sub-harmonic series includes the notes C$_3$ (sub-harmonic 2 of frequency $\frac{1}{2}\fff$), F$_2$ (sub-harmonic 3 of  frequency $\frac{1}{3}\fff$), C$_2$ (sub-harmonic 4 of frequency $\frac{1}{4}\fff$), A$\flat_1$ (sub-harmonic 5 of frequency $\frac{1}{5}\fff$), etc. (see Table \ref{f-under-over}).
The crucial point is that  {\em the first five sub-harmonics generate an inversion of a minor triad} (in case of the C$_4$, an F minor chord: A$\flat_1$–C$_2$–F$_2$).\footnote{To retrieve the root of the F minor triad, the sixth sub-harmonic must be added. Adding further sub-harmonics blurs the root again. Note that minor triad notes also appear in the overtone series, though much later and mixed with other harmonics unrelated to the minor chord. For instance, the 6th, 7th, and 9th harmonics of  C$_4$ correspond to a G minor chord (G$_6$–B$\flat_6$–D$_7$). Beside the fact that the B$\flat_6$ is too flat (with respect to the natural or tempered scale), it is hard to perceive this triad due to the strong presence of lower frequency components C and E.}

This interpretation remained a matter of controversy for several centuries \dario{\cite{rameau1722,Hauptmann1853,Oettingen1866,riemann1893,lewin1982,Levy1985}},
including the question of whether sub-harmonics are actually perceived,
as well as the search for sub-harmonics in brass and string instruments 
played in a nonconventional way  \citep{abbado1964,abbado1965,dabbene1973}. See also \citep{pamcutt1989harmony,wienpahl1959,Daipra2022}. 

\begin{widetext}

  \begin{center}
    \captionof{table}{The harmonic and sub-harmonic series corresponding to C$_4$.}
    \label{f-under-over}

    \begin{tabular}{ccccccccccccccc}
      \hline
      $\ldots$      & A$\flat_1$ & C$_2$    & F$_2$      & C$_3$    & C$_4$ & C$_5$    & G$_5$ &
      C$_6$         & E$_6$      & G$_6$    & B$\flat_6$ & C$_7$    & D$_7$ & $\ldots$           \\
                    & $\fff/5$   & $\fff/4$ & $\fff/3$   & $\fff/2$ &
      $\fff=262$ Hz & $2\fff$    & $3\fff$  & $4\fff$    & $5\fff$  &
      $6\fff$       & $7\fff$    & $8\fff$  & $9\fff$    &                                       \\
      \hline
    \end{tabular}
  \end{center}

\end{widetext}

Even though Zarlino and contemporaries hypothesize the existence of the sub-harmonic series, they could not know -- due to
the limited understanding of physics at that time -- that the sub-harmonic series is not physically present in the original harmonic sound:
mathematically, no Fourier components exist below the fundamental frequency.
This, however, does not mean that Zarlino and contemporaries were forcibly wrong, as the sub-harmonic series could be generated by the intricate structure of the inner ear or by the auditory cortex itself.
The question of whether the perception of the sub-harmonic series is real or not remains an open question in psychoacoustics \citep{pamcutt1989harmony}.

It is worth noting that the sub-harmonic series is often associated with the perception of Tartini's third tone (or combination tone), which refers to the perceived frequency $\fff_2 - \fff_1$ (or more in general $|n_1\fff_1-n_2\fff_2|$ with $n_1$ and $n_2$ integer)
when two  harmonic sounds of frequencies $\fff_2>\fff_1$ are played simultaneously. When $\fff_1$ is very close to $\fff_2$, Tartini's third sound is simply a {\em beat}. Similarly to  the sub-harmonic series, Tartini’s third tone is not physically present (see \citep{Moore2012,caselli2018} and references therin).
, it has been widely believed that Tartini's third tone is caused by nonlinear phenomena present in the inner ear.
\footnote{Several theories trying to explain the combination tone have been put forward over the centuries: Tartini's theory \citep{Tartini1754,Tartini1767}, Lagrange's theory \citep{Lagrange1759,Lagrange1762} and -- the most widely accepted -- Helmholtz's theory \citep{Helmholtz1863}.  See, for instance,  \citep{Moore2012,caselli2018,abbado1965,Abbado1972TerzoQuartoSuono} for an historical perspective. Here, we just mention that in contrast to Helmholtz's theory which predicts $\fff_2-\fff_1$  (for $\fff_2>\fff_1$) as main combination tone, Lagrange's theory predicts, assuming $\fff_2/\fff_1$  rational, the greatest common divisor between $\fff_1$ and $\fff_2$.
  More recent theories, based on the nonlinear response introduced by the active amplification provided by the outer hair cells, predict the additional perception of the combination tone $2\fff_1-\fff_2$ for $\fff_2>\fff_1$. See below for more detail.}
More recent research suggests that such a tone may be generated in deeper brain regions, as well, since it  is still perceived when the two tones are delivered separately to each ear via headphones \citep{binaural_beats}.

In this paper, we show that both the sub-harmonic series and Tartini's third  tone $\fff_2-\fff_1$ can be explained within a linear cochlear framework based on vibrating strings, provided that the information transmitted from the cochlea to the auditory cortex is assumed to be the energy stored in the strings. The nonlinearity required to explain Tartini's third tone is then inherent to the quadratic nature of the energy.

\subsection{A linear model for the cochlea}

\begin{figure}
  \begin{center}
    ~~~~~~~\includegraphics[width=6.9truecm]{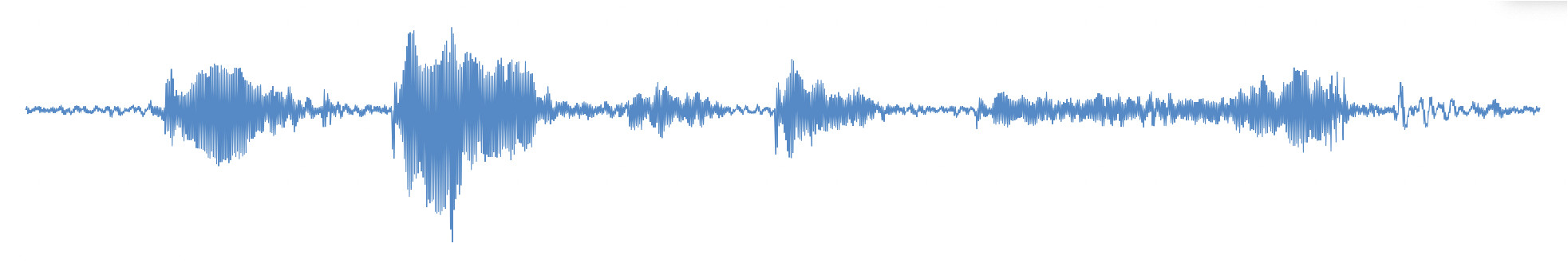}\\[2mm]\includegraphics[width=7.5truecm]{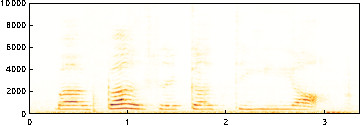}
    \caption{The sound (top) and its spectrogram (the ``sound image'',  bottom) \label{f-andrea}.}
  \end{center}
\end{figure}

The human auditory system is composed of the sensory organs, the ears (divided in inner, middle, and outer), which capture the physical sounds and convert them to neural signals that are then transmitted to the auditory cortex via the auditory nerve. In this paper, we focus on the cochlea,  a specific part of the inner ear,  which is the main organ responsible for sound perception.
Roughly speaking, the whole function of the outer and middle ear is to capture the incoming sound waves and transmit them to the cochlea, which is responsible for decomposing the sound into its fundamental frequencies, thereby constructing the so-called spectrogram (see Figure \ref{f-andrea}).
This spectrogram is then transmitted to the auditory cortex A1 via the auditory nerve, where further processing occurs.
It is worth to mention that the extant similarities between the auditory and visual cortices, have led some researchers to refer to the spectrogram as the ``sound image''. This conveys the idea that we do not hear sounds per se, but rather we ``see'' them (see, for instance, \citep{Moore2012,sacchelli}).

\begin{figure}
  \begin{center}
    \scalebox{0.87}{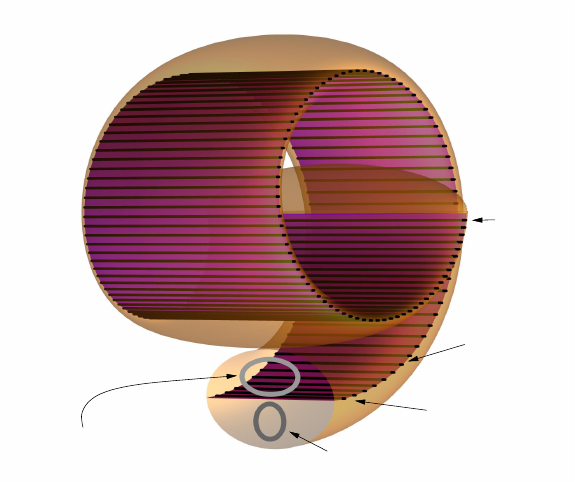}
    \caption{The Cochlea. \label{f-cochlea}}
  \end{center}
\end{figure}

The cochlea is a complex, fluid-filled structure containing a liquid called endolymph, which is set in motion by incoming sound transmitted through the outer and middle ear (pinna, tympanic membrane, and auditory ossicles). Mechanical vibrations reach the endolymph through the oval window, a membrane separating the middle ear from the cochlea that converts these vibrations into pressure waves within the fluid.
(See Figure \ref{f-cochlea}.)
In turn, these pressure waves act as forces on the {\em basilar membrane}, a thin structure extending along the length of the cochlea which is the main object of interest of this work.
This membrane has a varying width and stiffness along its length, being narrower and stiffer at the base (the entrance point of the cochlea) and wider and more flexible at the apex (the deepest point in the cochlea). This variation in mechanical properties allows different regions of the basilar membrane to resonate at different frequencies.   Such vibrations are detected by {\em inner hair cells}, which send signals roughly corresponding to the amplitude of the vibrations. (Additional hair cells, known as  {\it outer hair cells}, serve to amplify vibrations but will not be discussed here.\footnote{In this paper we
  consider a passive model of the cochlea. For models including the active action of  outer  hair cells, see, for instance, \citep{active1,active2} and references therein.})
In other words, the basilar membrane, with its hair cells,
performs a frequency analysis of the incoming sound.

The basilar membrane can be modeled as a collection of tensioned strings of varying lengths, linear mass, and tension.
We consider here the simple possible model in which the strings are {\bf noninteracting} and are modeled by a simple {\bf linear} damped {\bf string equation}.
Each string has its own resonance frequency (see below for the precise definition), and the oscillations are damped due to various mechanisms, the most immediate being the immersion in a viscous fluid. Strings closer to the entrance point of the cochlea (the base) have the highest resonance frequencies (around $20$KHz), strings closer to the deepest point in the cochlea (the apex) have the lowest resonance frequencies (around $20$Hz).

Parametrizing the different strings by $x\in[0,L]$ where $0$ is the base and $L$ the apex, we let $E(x,t)$ to denote the function encoding the information transmitted to the auditory cortex at time $t$ by the hair cells on the string at position $x$.
One crucial hypothesis we make in this paper is to assume that:

\begin{itemize}
  \item [{\bf (P) }]    the signal sent to the auditory cortex by the cochlea is, for each string $x$, the energy $E(x,t)$ stored in it.
\end{itemize}

When the input sound is harmonic, we perform a detailed analysis of the function $E(x,t)$ in the long-time limit (i.e., after the transient dynamics have decayed, see Sections \ref{s-sinusoidal} and \ref{s-hypotoni}).
Our main results can be summarized as follows:
\begin{enumerate}
  \item The function $E(x,t)$ is almost constant in time and reaches its maximum at the value of $x$  corresponding to the string  whose resonance frequency  is closest to the frequency of the incoming sound. This value corresponds to the main {\em peak} of the function. However, secondary -— though generally smaller -— peaks are also present. (See Figure \ref{f-arpa}.) Let us examine why.

        \begin{figure}[th]
          \begin{center}
            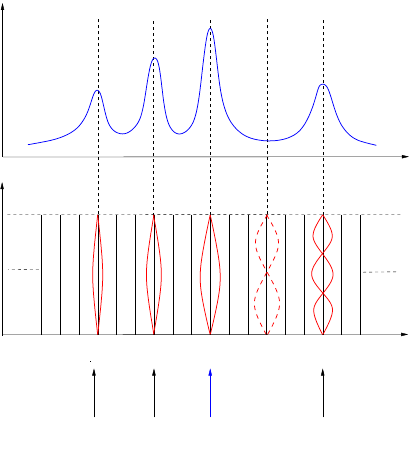
            \caption{
              The basilar membrane (bottom) is modeled as a set of vibrating strings responding to an incoming harmonic sound of frequency $\fff$. The signal, which contains several harmonics (e.g.,  the sawtooth of Figure \ref{f-dente-di-sega-Fourier}),  excites the strings whose resonance frequencies are $\fff$ (fundamental), $2\fff$ (second harmonic), $3\fff$ (third harmonic), and so on, as well as $\frac{\fff}{3}$ (third sub-harmonic), $\frac{\fff}{5}$ (fifth sub-harmonic), etc. As explained in Section \ref{s-matematica}, the sub-harmonics $\frac{\fff}{2}$, $\frac{\fff}{4}$, etc. are absent (the case $\frac{\fff}{2}$ is shown with a dashed line). The upper {\color{black} panel is a qualitative illustration of} the energy stored in the strings. Further details are given in Sections \ref{s-hypotoni} and \ref{s-misto}, {\color{black} in which we}  show that residual temporal oscillations occur and that sub-harmonics of the harmonics are also activated.}
            \label{f-arpa}
          \end{center}
        \end{figure}

        {\bf Peaks near  to the harmonics.}
        Assume that a harmonic sound corresponding to C$_4$ ($\fff = 262$Hz) is perceived.  What happens in reality is that, although the dominating frequency corresponds to a C$_4$, the sound contains several harmonic components of frequencies $2\fff$, $3\fff$, $4\fff$, etc.,  whose amplitudes and phases depend on the shape of the incoming periodic signal (see Figure \ref{f-dente-di-sega-Fourier} for an example). Each of these components will excite the string whose resonance frequency is close to its own frequency, producing peaks in $E(x,t)$ at the positions corresponding to frequencies $2\fff$, $3\fff$, $4\fff$, etc., thus
        enabling the perception of the timbre corresponding to the original harmonic sound.\\[1mm]

        {\bf Peaks near to the sub-harmonics.} While the previously examined peaks where due to the presence of harmonics in the original sound, other secondary peaks appear due to the strings themselves being sensitive to integer multiples of their resonance frequencies.  This is due to the fact that a string has several oscillation modes.
        More precisely, if the fundamental mode of a string corresponds to the frequency
        $\fff_0$, the next modes correspond to frequencies   2$\fff_0$, 3$\fff_0$ etc.
        Hence, if a harmonic sound   corresponding to the C$_4$  ($\fff=262$Hz)  is perceived, this will activate the first mode of the strings whose frequency of resonance  is  close to $\fff$ but also the second mode of the  strings whose  frequency of resonance   is    close to $\frac12\fff$, the third mode  of the  strings whose  frequency of resonance  is  close to $\frac13\fff$, etc...
        Notice that all these strings will vibrate at the frequency of the forcing signal ($\fff=262$Hz) but the information that is transmitted to the auditory cortex (depending only to the position of the activated string) is the presence  of oscillations at the frequencies of  $\frac12\fff$, $\frac13\fff$ etc. These frequencies correspond exactly to the frequency of the sub-harmonics, and this idea explains their perception.\footnote{ The same reasoning
          also applies  to all the harmonics of the incoming sound, producing sub-harmonics of the different harmonics. See Section \ref{s-misto}.}

        As we shall  see, in the simplest and arguably more natural model, certain symmetries imply that  {\bf only the odd sub-harmonics} appear (see Section \ref{s-matematica}). These are the most important ones since they contain the minor triad (the even sub-harmonics would  produce only lower octaves of the original sound).

        Notice that, contrary to harmonics, sub-harmonics are present even if the original sound is a pure sine wave. In fact, in the case of a pure sinusoid, they become even more evident, as there are no higher harmonics to mask their perception. This is could be one of the reasons why pure sine waves -- or, for example, the sound of a clarinet, which lacks all even-numbered harmonics -- are perceived as particularly “deep” or ``dark'' in timbre.

  \item The function $E(x,t)$ depends quadratically on the sound amplitude. This nonlinear behavior -- as we will see -- gives rise to interference phenomena that allows, e.g., to explain Tartini's third sound regardless of the linearity of the underlying model.

        More precisely,  as explained in Section \ref{s-tartini}, for a combination of two sinusoidal sounds of frequencies {\color{black} $\fff_2>\fff_1$}, the nonlinearity of the energy stored in each string $E(x,t)$ permits, as well, to see the appearance of residual oscillations in $E(x,t)$ of frequencies $2{\cal F}_1$, $2{\cal F}_2$, ${\cal F}_2-{\cal F}_1$, and ${\cal F}_1+{\cal F}_2$.
        Such oscillations in the energy of the strings are transmitted to the endolymph and returned back to the strings, producing the perception of the  combination tone (notice, however, that  such feedback is not described in our linear model).  We refer to Section \ref{s-tartini} for an  explanation of why the combination tone ${\cal F}_2-{\cal F}_1$ is more easily perceived than the others.
\end{enumerate}

\subsection{Connection with the literature}

The description of the cochlea as a continuum of linear vibrating strings dates back to \cite{Helmholtz1863} (see also \cite{reichenbach2014} and references therein). The main novelties of the present work are
  {\bf i)} the introduction of a model (actually non linear) for the information transmitted to the auditory cortex;
{\bf ii)} the consideration, in addition to the fundamental mode, of higher oscillation modes of the vibrating strings.

We should mention that a wide range of sophisticated models of the cochlea have been investigated, including nonlinear and active formulations (see, for instance, \citep{duifhuis2011modeling,RoblesRuggero2001,active1, active2} and references therein) as well as models focusing on traveling waves \citep{bekesy1947variation,von1960experiments,duke2003active,thesi}. In contrast to these approaches, our framework does not rely on the description of traveling waves, which is not particularly relevant for our purposes, since we are mainly concerned with periodic input signals and their steady-state response.

\subsection{Organization of the paper}

The paper is organized as follows In Section \ref{s-model} we introduce the model and describe its unforced oscillations, its response to a pure sinusoidal signal and  its response to a harmonic sound. In Section \ref{s-energy} we present our model for the signal sent to the auditory cortex, namely  the energy stored in the different strings. In Section \ref{s-hypotoni} we study the energy stored in the different strings under the action of a sinusoidal signal and show the emergence of sub-harmonics. In Section \ref{s-misto} we study the energy stored in the different strings under the action of a harmonic sound and show the presence of harmonics and sub-hamonics at the same time.  In Section \ref{s-tartini} we study the emergence of a combination tone when two sinusoidal signals of different frequencies are applied to the system.

Throughout the study,
whenever possible, we used experimentally validated parameters for the numerical simulation.
Unfortunately, it is very hard to obtain these parameters experimentally and their  knowledge  is fairly limited. We use two sets of parameters. The first set is extracted from \citep{nobili2003} and a second set is readapted so to obtain a more reasonable range of frequency covered by the model and to fit an assumption allowing to make certain computations more explicitly (see hypothesis {\bf (A03)} in Section  \ref{s-physics}).  These choices are explained in Appendix  \ref{s-parameters}.
Our model is, however, built in such a way that other values of the relevant parameters  can be easily used. The emergence of the sub-harmonic series  and of the combination tone appear  to be independent from the choice of parameters, at least in a reasonable range.

For the reader's convenience, we have collected all the notation used throughout the paper in Table \ref{tab:parameters}, at the end of the paper, along with a reference to its first appearance.

\section{A model of the basilar membrane as a set of noninteracting vibrating strings\label{s-model}}

\subsection{The physics}
\label{s-physics}

In the following we are going to model the basilar membrane as a set of non-interacting strings. The different strings
are parameterized via their longitudinal position $x\in[0,L]$, where $0$ denotes the base
and $L$ the apex.
Each string has its own length $\ell(x)$, linear mass $\rho(x)$, tension $T(x)$ and damping $\gamma(x)$.

Let $z\in[0,\ell(x)]$ be the coordinate along one fixed string $x$. Let $u^x(t,z)$ the displacement of the string $x$ at position $z$ at time $t$. The external force acting on the strings, representing the sound, will be indicated by $F(z,t)$. Its coupling with the different strings could in principle depend on $x$ and will be indicated with $c(x)$.

Each string is going to be described by the following damped string equation
\begin{align}
  \left\{\begin{array}{l}
           \rho(x) u^x_{tt}=T(x)u^x_{zz}-\gamma(x) u^x_t+c(x)F(t,z), \\
           u^x(t,0)=u^x(t,\ell(x))=0.
         \end{array}
  \right.
  \label{string}
\end{align}

For most of the results in this paper, it is not necessary to know the explicit expressions of $\rho(x),T(x),\gamma(x), c(x)$. However, it will be useful sometimes to have models for these functions, mainly for numerical simulations. According to \citep{reichenbach2014}, we have that (more details  are given in Section\ref{s-parameters} and Appendix \ref{a-parameters}):
\begin{itemize}
  \item $\ell(x)$ is an increasing function of $x$. However, its variation is not very large.  Along  the paper we will often consider that $\ell$ is constant.
  \item The linear mass of each string is varying as a function of $x$ with an exponential law $\rho(x)=A_\rho e^{k_\rho x}$, with $A_\rho, k_\rho>0$. Strings closer to the apex are much heavier.
  \item The tension of each string  is varying as a function of $x$ with a negative  exponential law $T(x)=A_T e^{-k_T x}$ with $A_T, k_T>0$. Strings closer to the apex have less tension.

  \item {\color{black} The damping of each string is varying as a function of $x$ with an exponential law $\gamma(x)=A_\gamma e^{k_\gamma x}$, with $A_\gamma>0$ and $k_\gamma\in{\bf R}$.  See below for a discussion on the sign of $k_\gamma$.}

  \item  The function $c(x)$ is a decreasing function of $x$ modeling the fact that the energy of the sound is already partially dissipated when the sound reaches the most internal part of the cochlea (large $x$). However, the decrease of $c(x)$ is not very relevant and we will often consider the case $c=1$.
\end{itemize}

In the following, any time we use  explicit expressions for $\rho(x),T(x),\gamma(x), c(x),\ell(x)$, we will refer to the following assumption:
\begin{itemize}
  \item [{\bf (A0)}] We assume that $\rho(x)=A_\rho e^{k_\rho x}$, $T(x)=A_T e^{-k_T x}$,  $\gamma(x)=A_\gamma e^{k_\gamma x}$, with $A_\rho,$  $A_T,$ $A_\gamma$, $k_\rho$, $k_T>0$, $k_\gamma\in\mathbf{R}$. We also assume that  $c$ and $\ell$ are positive  constants and we  normalize $c=1$.
\end{itemize}
In the literature one also finds the following further assumptions  in addition to {\bf (A0)}:
\begin{itemize}
  \item [{\bf (A01)}]  $k_\gamma=k_\rho=k_T$. see, for instance \citep{thesi} and references therein.

  \item [{\bf (A02)}]  $k_\gamma= \frac{k_T+k_\rho}{2}$ (to have that $\gamma$ is proportional to the inverse of the undamped frequency of the string defined below in \eqref{xi}). See, for instance, \citep{thesi} and references therein.

  \item[{\bf (A03)}] $k_\gamma=0$ (same damping for all the strings). This assumption, that we often make  in the following, permits to simplify certain analyses. It is justified by the fact that sometimes in the literature one finds $k_\gamma>0$  (meaning higher damping for lower frequencies, see, for instance, \citep{thesi,szalai2015nonlinear}) and somtines $k_\gamma<0$  (meaning lower damping for lower frequencies, see, for instance, \citep{nobili2003,ni2014modelling}).  The case $k_\gamma=0$ has also be considered in \citep{zweig1976cochlear}.
\end{itemize}

Let us define  the {\em undamped fundamental angular frequency}\footnote{In this formula $\xi(x)$ is  an angular frequency because it is measured in rad$/$sec. The frequency in Hertz is $\frac{\xi}{2\pi}$.} of each string (see Section \ref{s-unforced} for the justification of this formula)
\begin{equation}
  { \xi(x) := \sqrt{\frac{T(x)}{\rho(x)} }\frac{\pi}{\ell(x)}}.
  \label{xi}
\end{equation}

As in all realistic models, along the paper we will assume that  $\xi$ is a decreasing function of $x$. This permits us to parametrize solutions of \eqref{string} by $\xi$ which is actually the only important parameter in the following.

In particular, this is true under {\bf (A0)}, which implies
\begin{equation}
  \xi(x) \equalexpl{under (A0)}\sqrt{{A_T}/{A_\rho}} \,\frac{\pi}{\ell} e^{-\frac{k_T+k_\rho}{2} x}
  \label{xi-A0}
\end{equation}

\begin{remark} {\bf (Range of frequencies covered by the model)}
  \label{r-range}
  Let us re-write  \eqref{xi-A0} as
  \begin{align}
    \xi(x)         & =\tilde Ae^{-\tilde kx}, \label{mentecatto}                                                   \\
    \mbox{ where } & \tilde A= \sqrt{{A_T}/{A_\rho}} \,\frac{\pi}{\ell} \mbox{ and } \tilde k=\frac{k_T+k_\rho}{2}
  \end{align}
  The corresponding frequency range in Hz is {\color{black} $[\xi(L)/(2\pi),\xi(0)/(2\pi)]=[\tilde A e^{-\tilde k L}/(2\pi)     ,   \tilde A/(2\pi)   ]     $.}

\end{remark}

\subsection{The Mathematics}
\label{s-matematica}

Let us re-write \eqref{string} as
\begin{equation}
  u_{tt}=\frac{T}{\rho }u_{zz}- \frac{\gamma}{\rho} u_t+\frac{c}\rho F(t,z),~~~u(t,0)=u(t,\ell)=0,
  \label{ondacce}
\end{equation}
(here by simplicity of notation we have dropped the dependence on $x$ of $u,~\rho, ~T, ~\gamma,~c,~\ell$).

Let us take a regular enough forcing term. To fix the ideas let us take $F(t,\cdot) \in L^2([0,\ell])$ for a.e. $t\in{\bf R}$ and the map $t\mapsto \|F(t,\cdot)\|$ in $L^\infty({\bf R})$,
that is $F\in L^\infty({\bf R},L^2([0,\ell]))$.
We consider initial conditions $u(0,\cdot)\in H^1_0([0,\ell])$ and  $u_t(0,\cdot)\in L^2([0,\ell])$.
In this setting, the existence and uniqueness of solutions
$u\in L^\infty({\bf R},H^2([0,\ell])\cap H^1_0([0,\ell])$ is standard. See, for instance, \cite[Section 4.7]{Hale1988}.

Considering the basis  $e_n(z)=\sqrt{\frac2\ell}\sin(\frac{\pi}{\ell} n z)$ for $n=1,2,\ldots $ of $ L^2([0,\ell])\supset H^2([0,\ell])\cap H^1_0([0,\ell])$ we have
\begin{align}
  u(t,z) & =\sum_{n=1}^\infty p_n(t) e_n(z),~\mbox{ where }\nonumber                                                \\
  p_n(t) & = \langle e_n,u(t,z)\rangle =\int_0^\ell u(t,z) \sqrt{\frac2\ell}\sin(\frac{\pi}{\ell} n z) dz.\nonumber
\end{align}
This yields  the sequence of ODEs
\begin{equation}
  \ddot p_n=  \frac{T}{\rho } \left(-\frac{\pi^2}{\ell^2}n^2 \right)   p_n -\frac{\gamma}{\rho} \dot p_n+\frac{c}{\rho} f_n(t) ,~n=1,2,3,\ldots
  \label{pasta}
\end{equation}
where
$
  f_n(t)=\langle e_n,F(t,z)\rangle =\int_0^\ell F(t,z) \sqrt{\frac2\ell}\sin(\frac{\pi}{\ell} n z) dz.
$

Recall that the force $F(t,z)$
is applied to the string through endolymph motion driven by the auditory ossicles. Neglecting boundary effects in the vestibular canal, the endolymph motion depends only on $x$. Thus, we assume:
\begin{itemize}
  \item [{\bf (A1)}]  The applied force is independent of  $z$, i.e., $F(t,z)=F(t)$.
\end{itemize}
As we will see, this assumption will be responsible for the absence of even sub-harmonics.
Under {\bf (A1)} we have that
\begin{align}
  f_n(t) & = F(t) \int_0^\ell  \sqrt{\frac2\ell}\sin(\frac{\pi}{\ell} n z) dz\nonumber \\
         & =\left\{ \begin{array}{l}
                      F(t)  \sqrt{\frac{2}{\ell}}\frac{2\ell}{\pi} \frac{1}{n}=
                      F(t)   \frac{2\sqrt{2}}{\pi}  \sqrt{\ell}\frac{1}{n}
                      \mbox{~if }n\mbox{ is odd,} \\[1em]
                      0\mbox{~~~~ if }n\mbox{ is even.}\end{array}\right. \nonumber
\end{align}
Then \eqref{pasta} becomes
\begin{equation}
  \ddot p_n\!\equalexpl{under (A1)}\!-\xi_n(x)^2 p_n -\mu(x) \dot p_n+\nu_n(x) F(t),~n=1,2,3,\ldots\label{HOnn}
\end{equation}
where
\begin{align}
  \xi_n(x)                 & = \sqrt{\frac{T(x)}{\rho(x)} }\frac{\pi}{\ell(x)} n=\xi(x) n\nonumber \\
  {\mu(x)}                 & = \frac{\gamma(x)}{\rho(x)}\label{mum}                                \\
  {\color{black} \nu_n(x)} & =
  \left\{\begin{array}{l}
           \frac{c(x)}{\rho(x)}\frac{2\sqrt{2}}{\pi}  \sqrt{\ell(x)}\frac{1}{n}
           \mbox{~~~~ if }n\mbox{ is odd} \\0 \mbox{~~~~ if }n\mbox{ is even}
         \end{array}\right.\label{nonnepossopiu}
\end{align}

\subsubsection{Parametrization via $\xi$}
If we can invert  \eqref{xi}, we can re-write \eqref{HOnn} in function of $\xi$
\begin{align}
  \ddot p_n & \equalexpl{under (A1)}-\xi^2 n^2 p_n -\mu(x(\xi)) \dot p_n+ {\color{black}  \nu_n(x(\xi)) }F(t), \label{HOn2} \\
  n         & =1,2,3,\ldots \nonumber
\end{align}
Under {\bf (A0)} this can be done explicitly  and we obtain $x(\xi)=\frac{2}{k_T+k_\rho}\log\left( \frac\pi\ell \sqrt{\frac{A_T}{A_\rho}}        \frac1\xi\right)$.
As a consequence
\ffoot{old formula with $\gamma$ constant $ \mu(x(\xi)) \equalexpl{under (A0)}\frac\gamma{A_\rho}   \left( \sqrt{\frac{A_\rho}{A_T}} \frac{\ell }{\pi }   \xi  \right)^{\frac{2k_\rho}{k_T+k_\rho}}=:B_\mu \xi^\alpha\label{Amu}$}
\begin{align}
  \mu(x(\xi)) & \equalexpl{under (A0)}\frac{A_\gamma}{A_\rho}   \left( \sqrt{\frac{A_\rho}{A_T}} \frac{\ell }{\pi }   \xi  \right)^{\frac{2(k_\rho-k_\gamma)}{k_\rho+k_T}}=:B_\mu \xi^\alpha\label{Bmu}, \\
              & \mbox{ where } \alpha:={\frac{2(k_\rho-k_\gamma)}{k_\rho+k_T}}\nonumber
\end{align}
{
\begin{align}
  \!\!\! \nu_n(x(\xi)) & \equalexpl{under~(A0)}~~~~~~
  \left\{\begin{array}{l}   \frac1{A_\rho}   \left( \sqrt{\frac{A_\rho}{A_T}} \frac{\ell }{\pi }   \xi  \right)^{\frac{2k_\rho}{k_\rho+k_T}}     \frac{2\sqrt{2}}{\pi}  \sqrt{\ell} \frac1n \\  ~~~~=:B_\nu  \xi^{\frac{2k_\rho}{k_\rho+k_T}} \frac1n  \mbox{~ if }n\mbox{ is odd}\\ \\ 0 \mbox{~~~~~~~~~~~~~~~~~~~~~~ if }n\mbox{ is even}
         \end{array}\right. \label{Bnu}
\end{align}
}

Next  we will also need
\begin{equation}
  T(x(\xi))\equalexpl{under (A0)}A_T   \left( \sqrt{\frac{A_\rho}{A_T}} \frac{\ell }{\pi }   \xi  \right)^{\frac{2k_T}{k_\rho+k_T}} =:  B_T\xi^{\frac{2k_T}{k_\rho+k_T}}
  \label{TTT}
\end{equation}

In the following, with abuse of notation, we write $\mu(\xi)$,  {\color{black}  $\nu_n(\xi)$}  and $T(\xi)$ in place of $\mu(x(\xi))$, $\nu(x(\xi))$ and $T(x(\xi))$.\\[2mm]

Notice that
\begin{itemize}
  \item Under {\bf (A01)} we have $\alpha=0$ and $\mu(\xi)=B_\mu$, ~ $T(\xi)=B_T\xi$, ~  {\color{black}  $\nu_n(\xi)=B_\nu \frac \xi n$ (for $n$ odd and zero otherwise)}.
  \item Under {\bf (A02)} we have $\alpha=\frac{k_\rho-k_T}{k_\rho+k_T}\in (-1,1)$.
  \item Under {\bf (A03)} we have $\alpha\in (0,2)$ and $\mu(\xi)=B_\mu\xi^\alpha$, ~$T(\xi)= B_T\xi^{\frac{2k_T}{k_\rho+k_T}}$,  ~{\color{black}  $\nu_n(\xi)=B_\nu \frac{\xi^\alpha}{n}$ (for $n$ odd and zero otherwise)}.
\end{itemize}

\begin{remark}
  For the undamped case (i.e., $\xi>\mu(\xi)/2$) to occur at high frequencies,
  it is natural to assume  that $\alpha < 1$ (see Section \ref{s-unforced}).
  This is the case for the set of parameters given in Appendix \ref{s-parameters}.
\end{remark}

\subsubsection{Unforced case}
\label{s-unforced}

Let us discuss briefly the solution of the equation \eqref{HOn2} in the case in which there is no forcing term ($F(t)=0$), i.e.
\begin{equation}
  { \ddot p_n= -\xi^2 n^2 p_n -\mu(\xi) \dot p_n~~~n=1,2,3,\ldots}
  \nonumber
\end{equation}
This can be alternatively written as a first order equation

\begin{equation}
  \left(
  \begin{array}{c}   \dot p_n \\   \dot q_n
    \end{array}
  \right)=M_n   \left(
  \begin{array}{c}   p_n \\   q_n
    \end{array}
  \right)
  \mbox{ where } M_n=  \left(
  \begin{array}{cc}
      0     & n\xi      \\
      -n\xi & -\mu(\xi)
    \end{array}
  \right),\nonumber
\end{equation}

\begin{itemize}
  \item If there is no damping $\mu(\xi)=0$, the solutions are oscillations of angular frequency $n\xi$. This is why $\xi$  has been called the undamped fundamental angular frequency of the string.
\end{itemize}
In the presence of damping, since the eigenvalues of the matrix  $M_n$ are
\begin{equation}
  -\frac{\mu(\xi)}{2}\pm i\sqrt{n^2\xi^2-\Big( \frac{\mu(\xi)}{2}\Big)^2},
  \label{autovalori}
\end{equation}
we have the following cases:

\begin{itemize}
  \item  \emph{Underdamped case:} $\xi>\frac{\mu(\xi)}{2}$ then for every $n$ the solutions are oscillations of angular frequency $\omega_n=\sqrt{n^2\xi^2-\Big( \frac{\mu(\xi)}{2}\Big)^2}$ and of exponentially decreasing amplitude. We call $\omega_n$ the $n$-resonance angular frequency of the string.

  \item  \emph{Overdamped case:} $\xi<\frac{\mu(\xi)}{2}$ then the above is true for $n$ large enough only. For $n\leq\frac{\mu(\xi)}{2\xi} $ the solutions go to zero without oscillating.

\end{itemize}

\subsubsection{Asymptotic response to a sinusoidal signal\label{s-sinusoidal}}

In this section, we apply the simplest possible sound (i.e., a sinusoidal function) to our string equation \eqref{HOn2}, and we study the corresponding solution for large time, once the solution has become periodic. Studying this case permits to see the emergence of the sub-harmonic series.

To this purpose,  we start directly from the decomposition in Fourier modes under {\bf (A1)}
\begin{equation}
  { \ddot p_n \equalexpl{under (A1)} -\xi^2 n^2 p_n -\mu(\xi) \dot p_n+ {\color{black}  \nu_n(\xi) }F(t) ,~~~n=1,2,3\ldots}
  \label{HOn}
\end{equation}
with
\begin{equation}F(t)=\sin(\kk t+\varphi )\label{poppa}
\end{equation}
Remarkably for every initial condition $p_n(t_0)=a$,  $\dot p_n(t_0)=b$ the solution of \eqref{HOn} converges  exponentially to the periodic solution (see e.g., \citep{Hale1988})
\begin{align}
   & p_n(t)~~~~~~~~ \equalexpl{\hskip -.9cm $F(t)=\sin(\kk t+\varphi )$}\nonumber \\
   & -  \nu_n(\xi) \frac{ \left(\left(\kk^2-\xi ^2 n^2\right) \sin (\kk
    t+\varphi)+\kk \mu(\xi)  \cos (\kk t +\varphi)\right)}{\kk^4+\kk^2 \left(\mu(\xi)
  ^2-2 \xi ^2n^2\right)+\xi ^4 n^4},\nonumber                                     \\
   & \mbox{ (here $n=1,2,3\ldots$).} \nonumber
\end{align}
In other words, this solution is globally attractive and periodic, and considering it amounts to ignoring the transient in the cochlear response.

  {\color{black} Notice that due to \eqref{nonnepossopiu}, we have that $p_n(t)=0$ for $n$ even.}
We can write the {previous formula} in the form\footnote{Using the fact that $\sin(\alpha) \cos(\kk t)+\cos(\alpha)\sin(\kk t)=\sin(\alpha+kt)\Rightarrow A\sin(\alpha) \cos(\kk t)+A\cos(\alpha)\sin(\kk t)=A\sin(\alpha+kt)$ and writing $A\cos(\alpha)=k^2-\xi_n^2$ and $A\sin(\alpha)=k\mu$.}
\begin{equation}
  { p_n(t) \equalexpl{$F(t)=\sin(\kk t+\varphi )$}{\cal R}_n(\xi,\kk)\sin(\kk t +\varphi+\phi_n(\xi,\kk)),~n=1,2,3\ldots}
  \label{sol-seno}
\end{equation}
where
  {
    \begin{align}
      {\cal R}_n(\xi,\kk) & =  \nu_n(\xi) \frac{1}{\sqrt{   \kk^4+\kk^2 \left(\mu(\xi)
      ^2-2 \xi ^2n^2\right)+\xi ^4 n^4 }}\label{RRn}                                   \\
      \phi_n(\xi,\kk)     & =\arctan\frac{\kk \mu(\xi)}{\kk^2-n^2 \xi^2}\label{PHIn}
    \end{align}
  }

\begin{remark}
  Under forcing, all strings oscillate with angular frequency $\kk$, even those ones for which we are in the overdamped case.
\end{remark}

\begin{remark}  The function ${\cal R}_n$  is always well defined for every (positive) value of $n,\xi,\kk$ and every function $\nu(\xi)$, $\mu(\xi)$.\end{remark}

If, in addition to $F(t)=\sin(\kk t+\varphi)$, we assume {\bf (A0)}, we have
\begin{align}
   & {\cal R}_n(\xi,\kk) ~~~\equalexpl{\!\!\!\!\!under (A0)}\nonumber \\
   & \left\{
  \begin{array}{ll}
    \frac{{\color{black}B_\nu \xi^{\frac{2k_\rho}{k_\rho+k_T}} }}{n}  \frac{1}{\sqrt{   \kk^4+\kk^2 \left( (B_\mu \xi^\alpha)
    ^2-2 \xi ^2n^2\right)+\xi ^4 n^4 }} ~~~ & n=1,3,5,\ldots \\
    0~~~                                    & n=2,4,6,\ldots\end{array}\right.
  \label{RN-A0}
\end{align}

\noindent
{\bf Study of  ${\cal R}_n(\xi,\kk)$ for $n$ odd}

\begin{itemize}
  \item For fixed $\xi$ its maximum is at $\kk_n^{\mathrm{max}}=\sqrt{n^2\xi^2-\frac{\mu(\xi)^2}2}$. Notice that this is not the $n$-resonance angular frequency of the string, which is
        $\omega_n=\sqrt{n^2\xi^2-\Big(     \frac{\mu(\xi)}2    \Big)^2}$

  \item For fixed $\kk$,  its maximum cannot be computed explicitly without an explicit expression of $\mu(\xi)$. However, we have the remarkable result that under {\bf (A03)} its maximum  is at
        $${ \xi_n^{\mathrm{max}}\equalexpl{under (A03)}\kk/n.}$$

        If we assume {\bf (A01)} we have that the maximum is a bit displaced:

        $${ \xi_n^{\mathrm{max}}\equalexpl{under (A01)}\frac{\sqrt{\kk} \sqrt[4]{B_\mu^2+\kk^2}}{n}.}$$

\end{itemize}

\begin{remark}  Notice that each string has three different frequencies associated with it:
  \begin{itemize}
    \item Its undamped resonance angular frequency $\xi$.
    \item Its resonance angular frequency  $\omega_1=\sqrt{\xi^2-\Big(     \frac{\mu(\xi)}2    \Big)^2}$.
    \item The angular frequency of an external signal provoking the largest oscillations  $\kk_1^{\mathrm{max}}=\sqrt{\xi^2-\frac{\mu(\xi)^2}2}$.

  \end{itemize}
\end{remark}

\begin{figure}
  \begin{center}
    \includegraphics[width=6.5truecm]{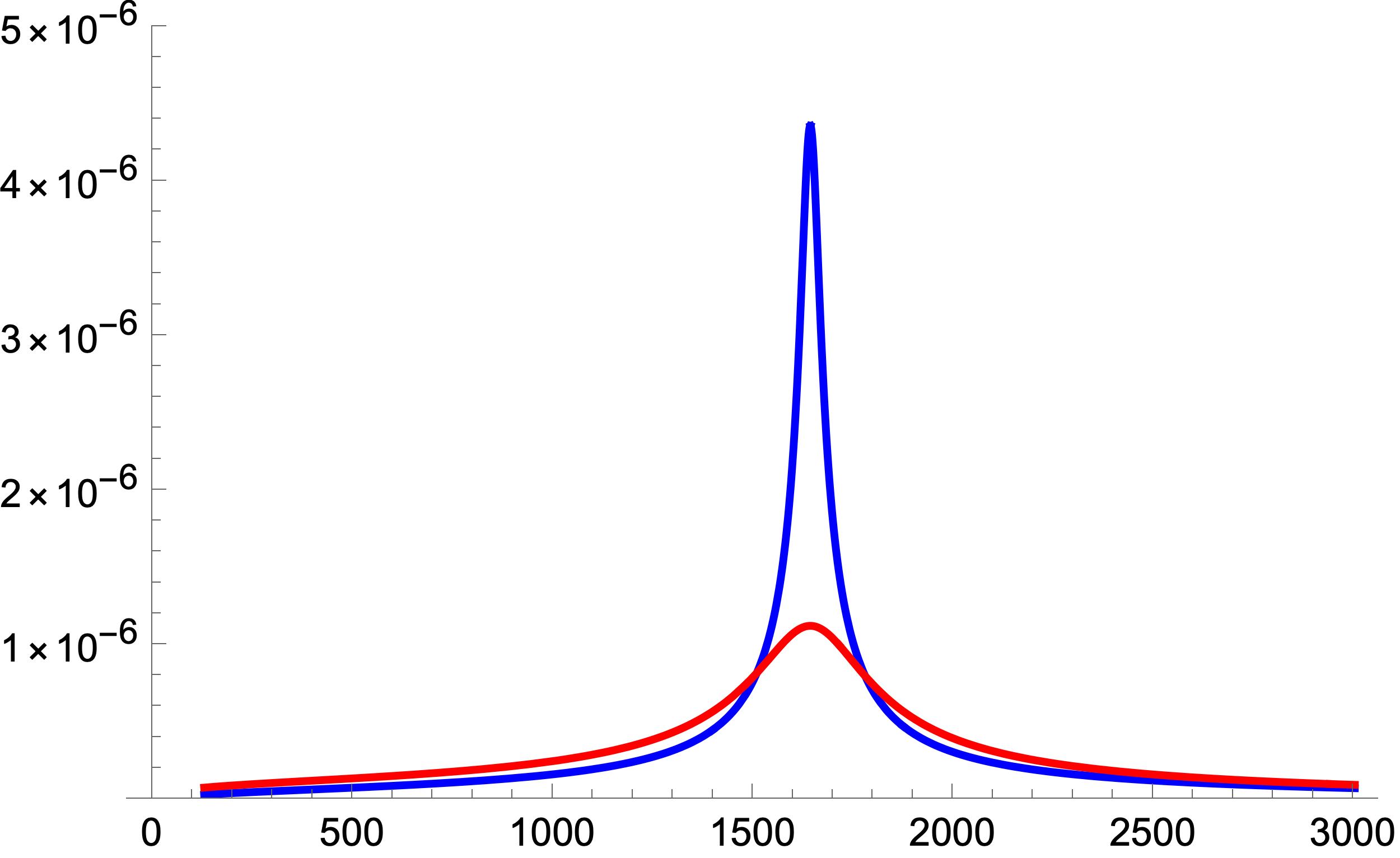}
    \caption{${\cal R}_1$ as function of $\xi$ for a sinusoidal input signal corresponding to C$_4$ (i.e. for $\kk$ fixed to $2\pi 262$ sec$^{-1}$, $\varphi=0$), using the parameter given in Appendix~\ref{s-parameters}, Table~\ref{t-parameters-original}. The blue curve corresponds to the parameters extracted from \cite{nobili2003}, while the red curve corresponds to the modified parameters.
      \label{f-R1}}
  \end{center}
\end{figure}

In Figure \ref{f-R1}, we depict ${\cal R}_1(\xi, 2\pi 262)$ -- i.e., we consider a sinusoidal input signal that corresponds to C$_4$ -- for the set of  parameters of Section \ref{s-parameters-original}
(parameters of \citep{nobili2003}) and of Section  \ref{SparametersModifA03}
(modified parameters and assumption {\bf (A03)}). Notice the similar qualitative behavior.

\subsubsection{Asymptotic response to a harmonic sound\label{s-periodic}}

Consider now a periodic signal $F(t)$ with angular frequency $\kk>0$.  {\color{black} Under the hypothesis given in Section \ref{s-matematica} we have that that
$F |_{ [0, \frac{2\pi}{\kk} ] }\in L^2[0, \frac{2\pi}{\kk}]$.} Then  $F(t)$ can be decomposed in its Fourier components
$$
  F(t)= \sum_{j=1}^\infty c_j\sin ( j\kk t +\varphi_j)
$$
(here for simplicity we are assuming that $F(t)$ has zero average, so that $j$ starts from 1).

By linearity, the corresponding response is (for $n=1,2,3,\ldots$)
\begin{equation}
  { p_n(t)=  \sum_{j=1}^\infty c_j  {\cal R}_n(\xi,j \kk)\sin(j \kk t +\varphi_j+\phi_n(\xi,j\kk)),}
  \label{sol-periodic}
\end{equation}
where  ${\cal R}_n$ and $\phi_n$, are given by \eqref{RRn} and  \eqref{PHIn}.

\section{A model for the information sent to the auditory cortex}
\label{s-energy}

The observed insensitivity to the relative phases of the Fourier components (see, for instance, \citep{Moore2012,ZwickerFastl1999}) is a strong indication that the signal transmitted from the cochlea to the auditory system is not the full solution $u^x(t,z)$ of equation \eqref{string}, but rather a smoothed version in which fast oscillations are averaged out.

For instance, in the case of the response to a sinusoidal input as discussed in Section \ref{s-sinusoidal}, what is sent from the cochlea is not the complete solution (here expressed through its  Fourier coefficients) $p_n(t)$ for $n=1,2,3,\ldots$, but rather a quantity only related to the amplitude of the oscillations ${\cal R}_n(\xi,\kk)$.

For a signal more complex than a pure sinusoid (e.g., the sum of two sinusoids with different frequencies), it is not immediately clear how to perform such averaging of the fast oscillations.

In the following, we introduce a natural assumption that accounts for at least part of this averaging process, and that leads to the emergence of the sub-harmonic series in a very natural way:

\begin{itemize}
  \item[{\bf (P)}] The signal transmitted from the cochlea to the auditory cortex is the energy stored in each string.
\end{itemize}
Referring to \eqref{string}, such energy is
\begin{equation}
  E(x,t)=  \int_0^{\ell(x)} \Big(  \frac12\rho(x) (u_t^x)^2+\frac12T(x)(u_z^x)^2 \Big)\,dz\label{energia-1}
\end{equation}
The first term is the kinetic energy due to the movement of the strings and  the second term is the potential (elastic) energy.

Decomposing in the Fourier basis used in Section \ref{s-sinusoidal} and writing $x$ in function of $\xi$ (with the usual abuse of notation) we obtain
\begin{align}
  E(\xi,t)= & \sum_{n=1}^\infty\frac12\rho(\xi) \Big(\dot p_n(t)^2+\xi_n^2 p_n(t)^2 \Big)\nonumber                                     \\
  =         & \frac12 \frac{\pi^2}{\ell(\xi)^2}T(\xi)\sum_{n=1}^\infty  n^2\Big(p_n(t)^2+\frac1{\xi^2 n^2}\dot p_n(t)^2 \Big)\nonumber \\
  =         & \sum_{n=1}^\infty E_n(\xi,t),\nonumber
\end{align}
where
\begin{align}
  E_n(\xi,t) & = \frac12 \frac{\pi^2}{\ell(\xi)^2}T(\xi) n^2\Big(p_n(t)^2+\frac1{\xi^2 n^2}\dot p_n(t)^2 \Big),\label{E-general}
\end{align}
is the energy stored at time $t$ in the Fourier mode $n$ by the string whose undamped resonance angular frequency is $\xi$.

\section{Perception of a sinusoidal signal and emergence of the sub-harmonic series}
\label{s-hypotoni}
In this section,  we are in the same framework as Section \ref{s-sinusoidal},
i.e., we assume $F(t,z)=\sin(\kk t+\varphi )$ and, in particular, we are working under {\bf (A1)}).
Using \eqref{sol-seno} we get that the signal transmitted by the cochlea to the auditory cortex is in this case
\begin{align}
  E(\xi,t)   & = \sum_{n=1,2,3,\ldots
  \ldots}E_n(\xi,t)\mbox{ where }\label{Eplutonio}                                                                        \\
  E_n(\xi,t) & \equalexpl{$\hskip -.7cm F(t)=\sin(\kk t+\varphi )$}
  ~~~~\frac12 \frac{\pi^2}{\ell(\xi)^2}T(\xi)   n^2  {\cal R}_n(\xi,\kk)^2 \times \nonumber                               \\
             & \left( 1+\left(\frac{\kk^2}{\xi^2 n^2} -1\right) \cos^2(\kk t+\varphi+\phi_n(\xi,\kk)) \right).  \nonumber
\end{align}

In the following, we call $E_n(\xi,t)$ the $n$th sub-harmonic.
\begin{remark}
  Note that the series \eqref{Eplutonio} is converging due to the term $n^4$ in formula \eqref{RRn}. Moreover, for fixed $n=1,2,\ldots$, $\kk,\xi>0$ ,  $E_n(\xi,t)$ is always positive.
\end{remark}

\begin{remark}
  Note that only the odd sub-harmonics are present. In addition, for $n$ odd, we have that  $E_n(\xi,t)$ is independent of time when $\xi =\kk/n$, i.e., when we are considering a string whose undamped resonance angular frequency
  (or one of its odd multiples)   coincides with the angular frequency of the external signal.
  Around the same value $\xi=\kk/n$
  the function  ${\cal R}_n(\xi,\kk)$ has a maximum.
  For $\xi \neq \kk/n$ the function $E_n(\xi,t)$ has residual oscillations with angular frequency $2\kk$.
  This is due to the fact that frequency  of $\cos^2(\kk t)$ is twice the frequency of
  $\cos(\kk t)$. The same occurs for every odd sub-harmonics, hence  the full series $E(\xi,t)$ oscillates with angular frequency $2\kk$.
\end{remark}

Summing over $n$, the presence of these maxima produces several peaks, one for every sub-harmonic, as explained in more detail next.

If we assume {\bf (A0)} and use \eqref{TTT} and \eqref{RN-A0}, we get
\begin{align}
  E(\xi,t)   & = \sum_{n=1,3,5,
  \ldots}E_n(\xi,t)\mbox{ where }\nonumber                                                                              \\
  E_n(\xi,t) & \equalexpl{$F(t)=\sin(\kk t+\varphi )$ and under {\bf (A0)} }
  \frac12 \frac{\pi^2}{\ell ^2}  B_T B_\nu^2 \xi^{2\frac{2k_\rho+k_T}{k_\rho+k_T}}  \times\nonumber                     \\
             & \frac{1}{{\kk^4+\kk^2 \left( (B_\mu \xi^\alpha)
  ^2-2 \xi ^2n^2\right)+\xi ^4 n^4 }}\times\nonumber                                                                    \\
             & \left( 1+\left(\frac{\kk^2}{\xi^2 n^2} -1\right) \cos^2(\kk t+\varphi+\phi_n(\xi,\kk)) \right).\nonumber
\end{align}

\begin{figure}
  \begin{center}
    \includegraphics[width=7.2truecm]{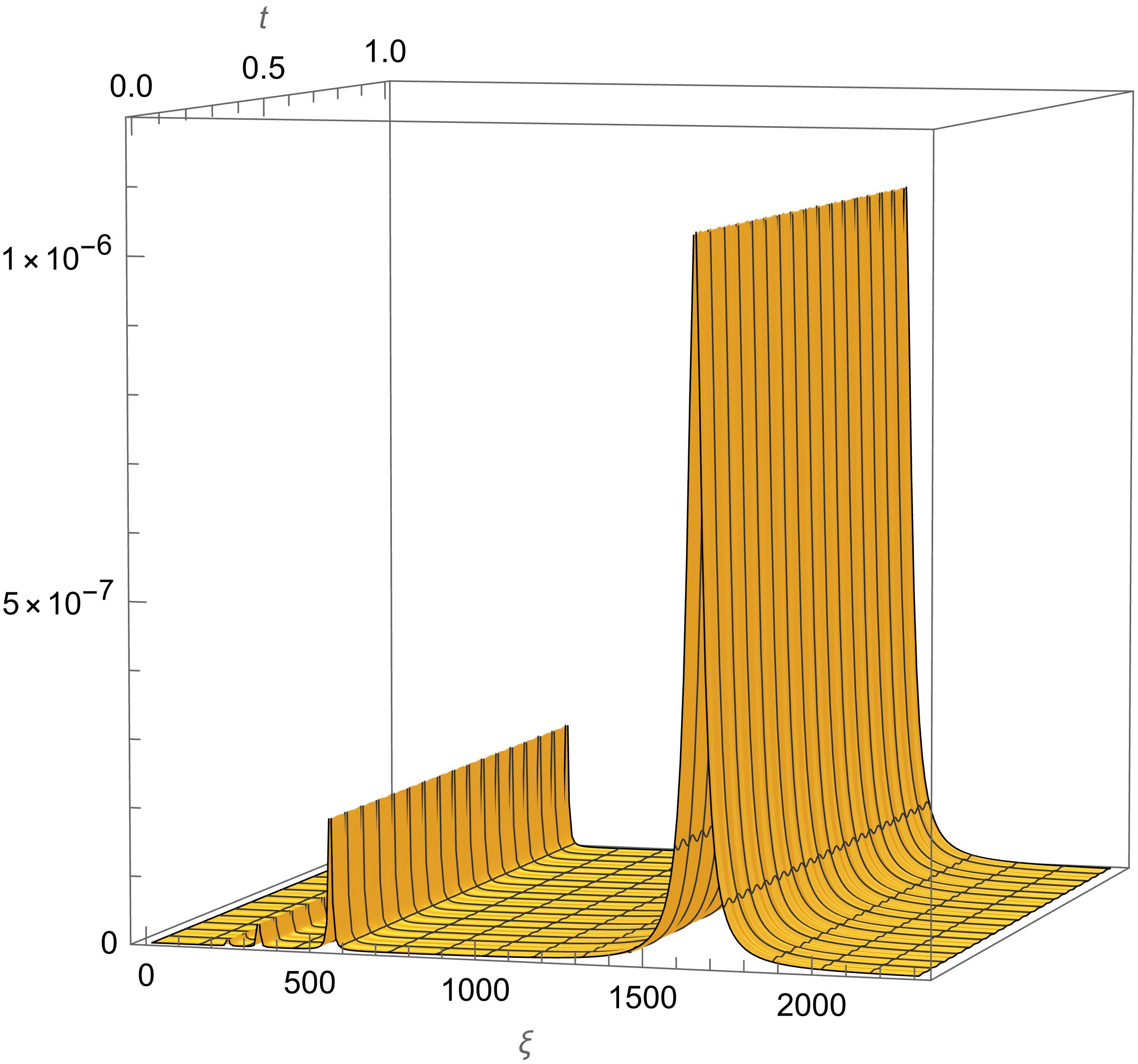}\\[2mm]\includegraphics[width=7.2truecm]{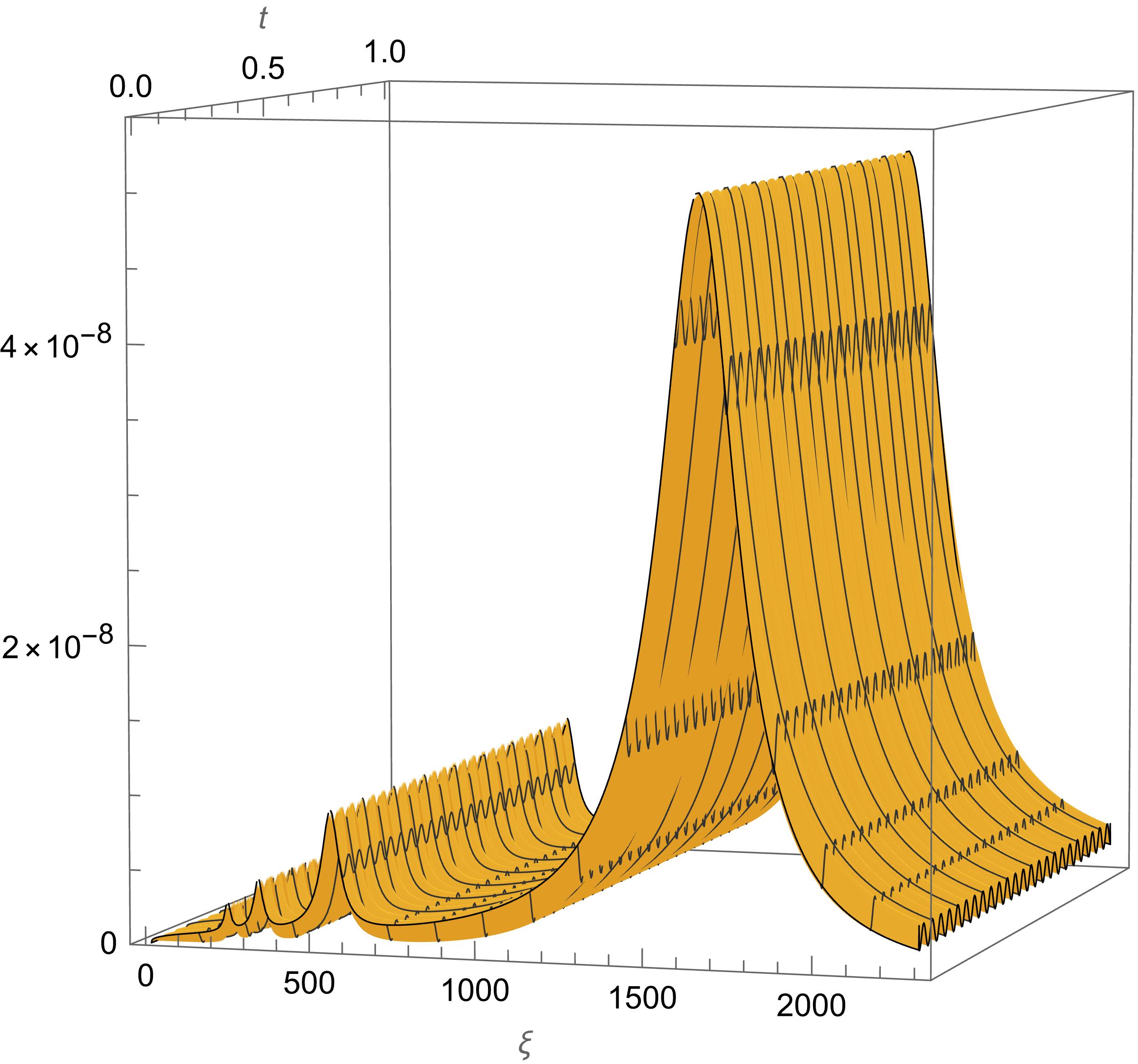}
    \caption{ {\color{black}  The energy stored in the strings $E(\xi,t)$ for
          an input that is a sinusoidal signal corresponding to the C$_4$ (i.e., for  $\kk=2\pi 262$ sec$^-1$, $\varphi=0$)  with the choice of parameters reported
          in Appendix~\ref{s-parameters}, Table~\ref{t-parameters-original}: third column (top) and fourth column (bottom).
          The higher peak corresponds to the frequency of the external signal. The lower peaks correspond to the sub-harmonics.
          \label{f-3D}}}
  \end{center}
\end{figure}

In Figure \ref{f-3D}, we depict the energy stored in the strings $E(\xi,t)$ for
an input that is a sinusoidal signal corresponding to the C$_4$.

\begin{figure}[th]
  \includegraphics[width=7truecm]{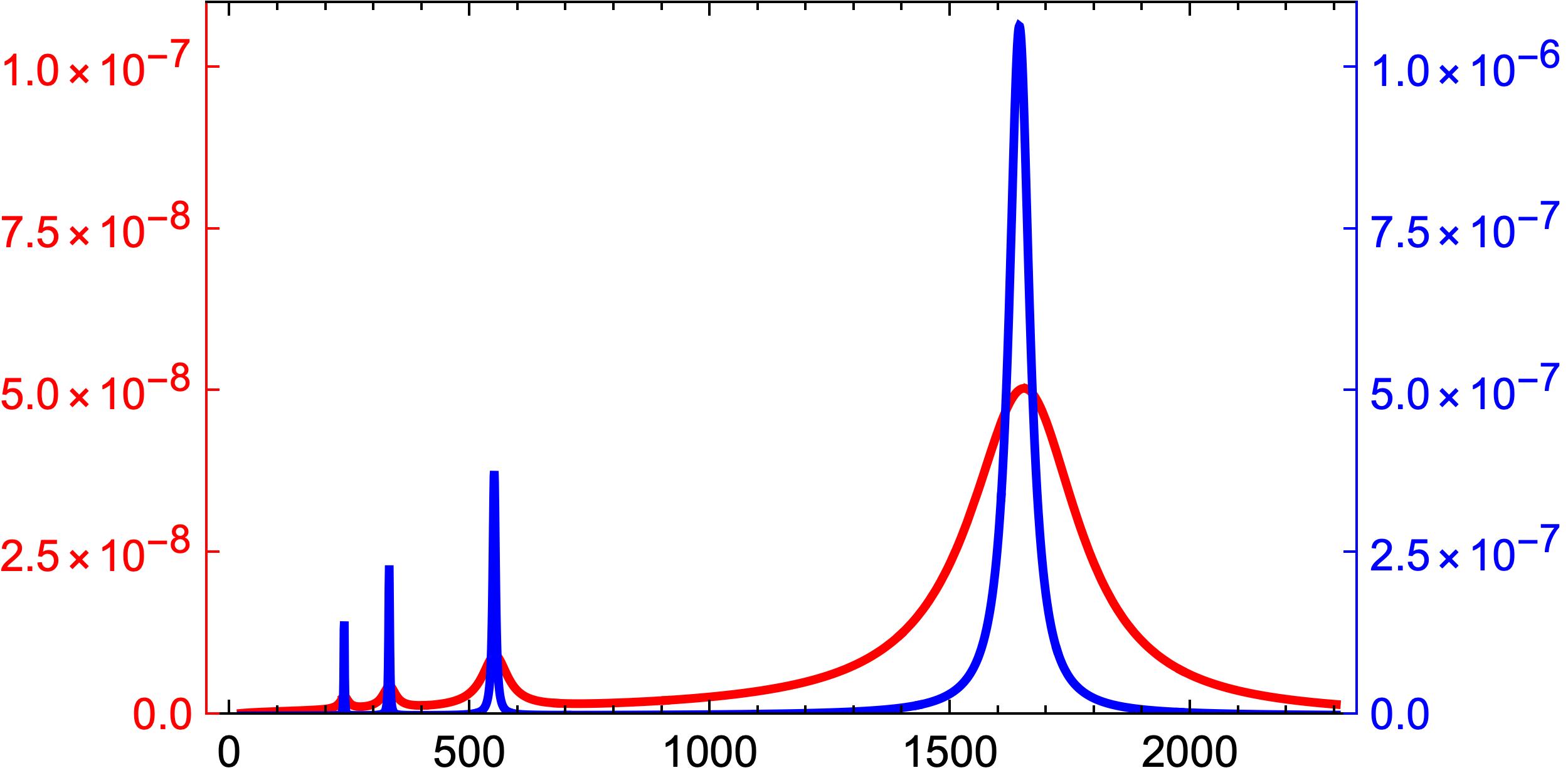}
  \caption{$E(\xi,t)$ for $t=0$ for the same parameters  of Figure  \ref{f-3D}. Parameter sets of Appendix  \ref{s-parameters}, Table~\ref{t-parameters-original}, third column for the blue function and fourth one for the red one.\label{f-NOt}}
\end{figure}

For better visibility of the sub-harmonic series, in Figure \ref{f-NOt} we draw $E(\xi,t)$ for $t=0$ (the variations in $t$ are much smaller than the variations in $\xi$).

\begin{figure}[th]
  \begin{center}
    \includegraphics[width=7truecm]{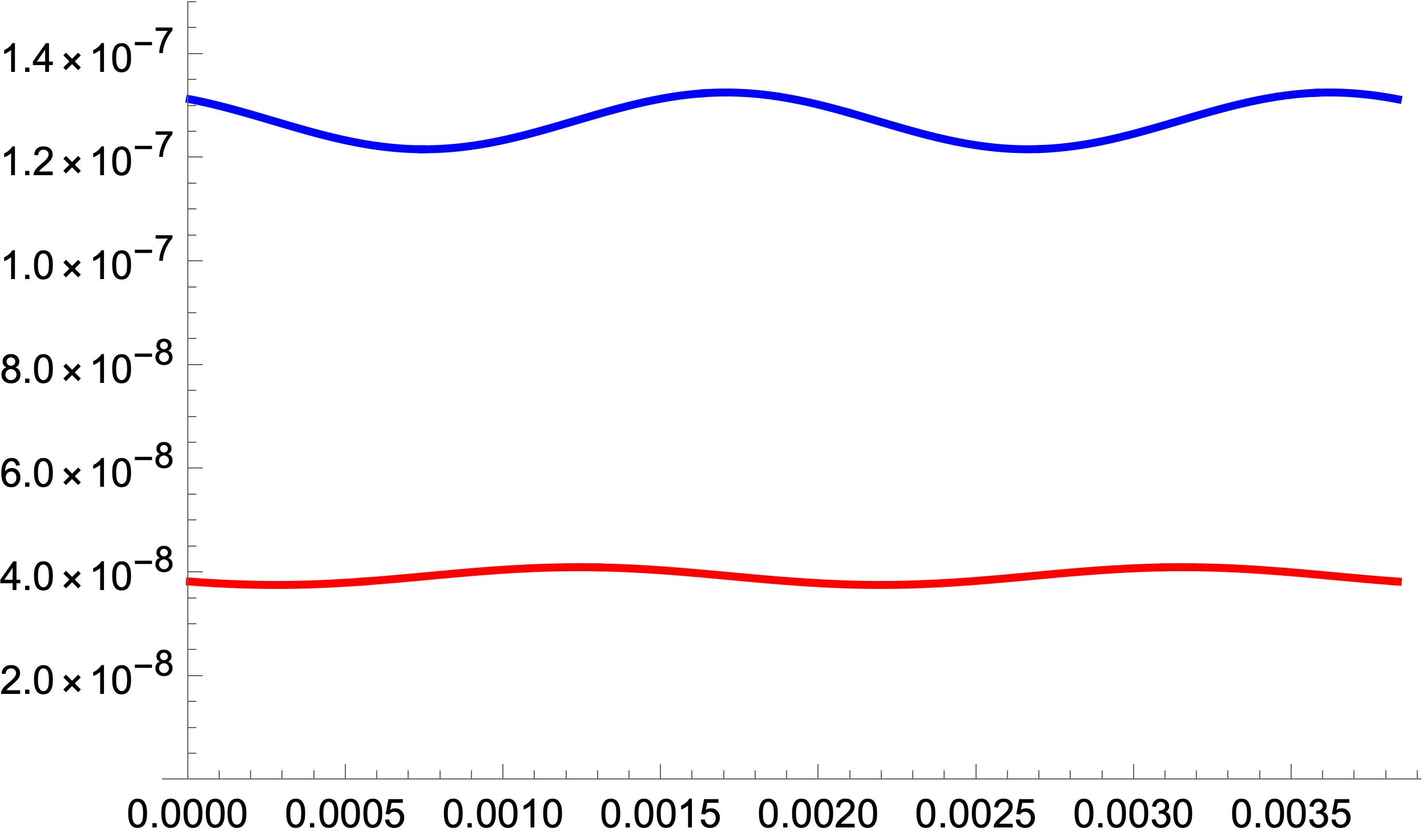}\
    \caption{
      For the same parameters as in Figure~\ref{f-3D}, we plot $E(\xi,t)$
      at $\xi=2\pi\times250\approx1570$ (close to, but not exactly at, the main
      peak) for $t\in[0,2\pi/\kk]$. The dependence on $t$ is periodic with
      angular frequency $2\kk$ and has small amplitude.
      \label{f-smallVt}
    }
  \end{center}
\end{figure}

To show that the variations in $t$ are small, in Figure \ref{f-smallVt}  we show $E(\xi,t)$ for fixed $\xi$.

\begin{remark}
  Notice that for fixed $\xi$, $E(\xi,t)$ is periodic with angular frequency $2\kk$.
  In the same spirit as what is explained in Section \ref{s-tartini}, this oscillation  is transmitted to the endolymph and then back to the forcing term. Although this phenomenon is not
  described by our linear model, it is going to  produce an additional peak at $2\kk$. This peak reinforces the perception of the second harmonic of the original sound.
\end{remark}

\subsection{Perception of a periodic signal}
\label{s-misto}

When $F(t)$ is a periodic signal, the corresponding energy
$E(\xi,t)=\sum_{n=1}^\infty E_n(\xi,t)$ is obtained by plugging \eqref{sol-periodic} into \eqref{E-general}.
Notice, however, that $E(\xi,t)$ is not a linear function of $p_n(t)$. As a result, the contributions of the different Fourier components of the original sound cannot be simply added; instead, they interact and give rise to interference phenomena. In other words, {\em sub-harmonics can be summed, but harmonics cannot}.
More details concerning this nonlinearity are given in Section \ref{s-tartini}.

In Figure \ref{f-completa} we draw $E(\xi,t)$  for a sawtooth corresponding to the C$_4$ for the parameters given in Appendix~\ref{s-parameters}. Notice the emergence of harmonics and sub-harmonics.

\begin{figure}
  \begin{center}
    \includegraphics[width=9truecm]{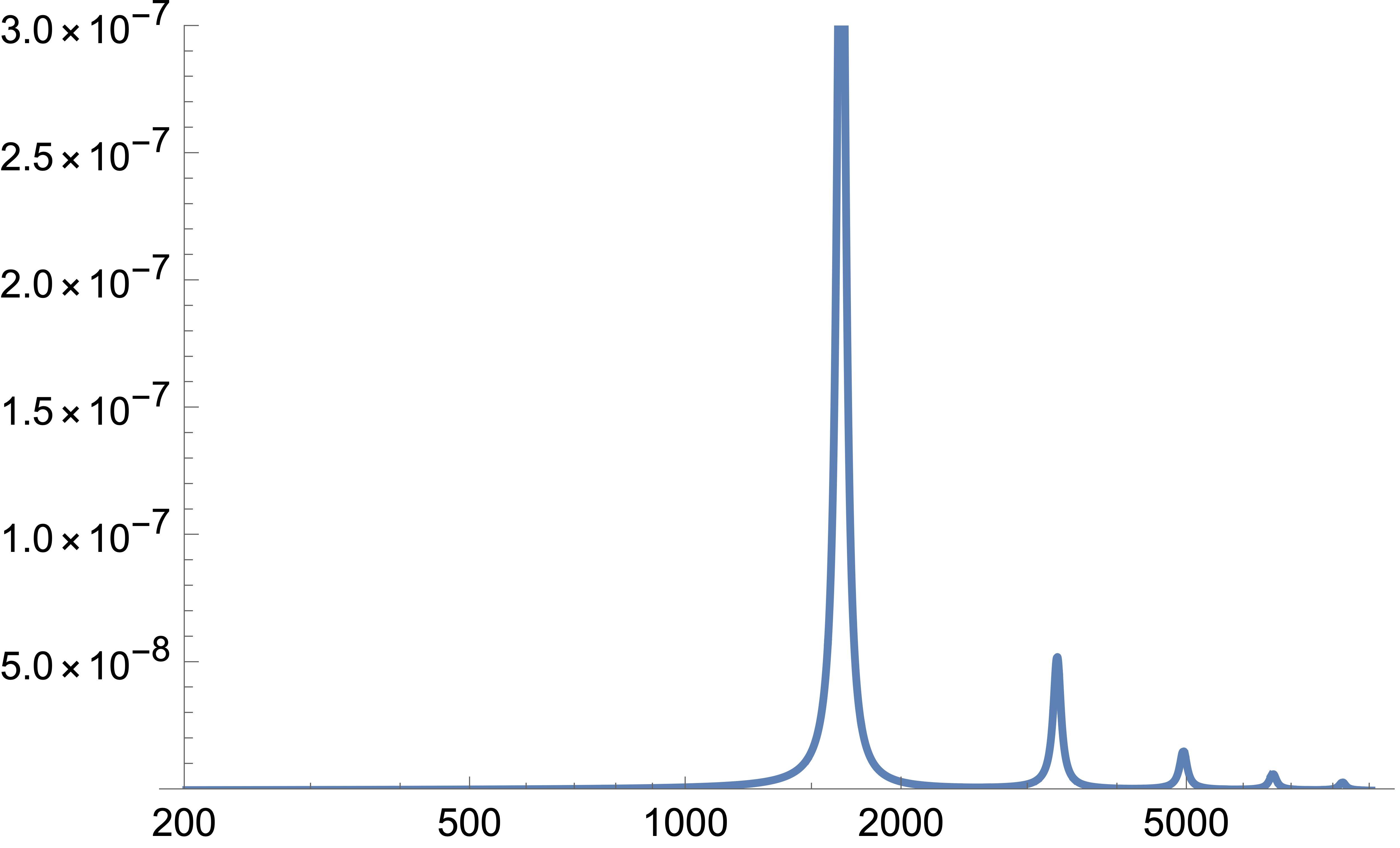}\\[2mm]\includegraphics[width=9truecm]{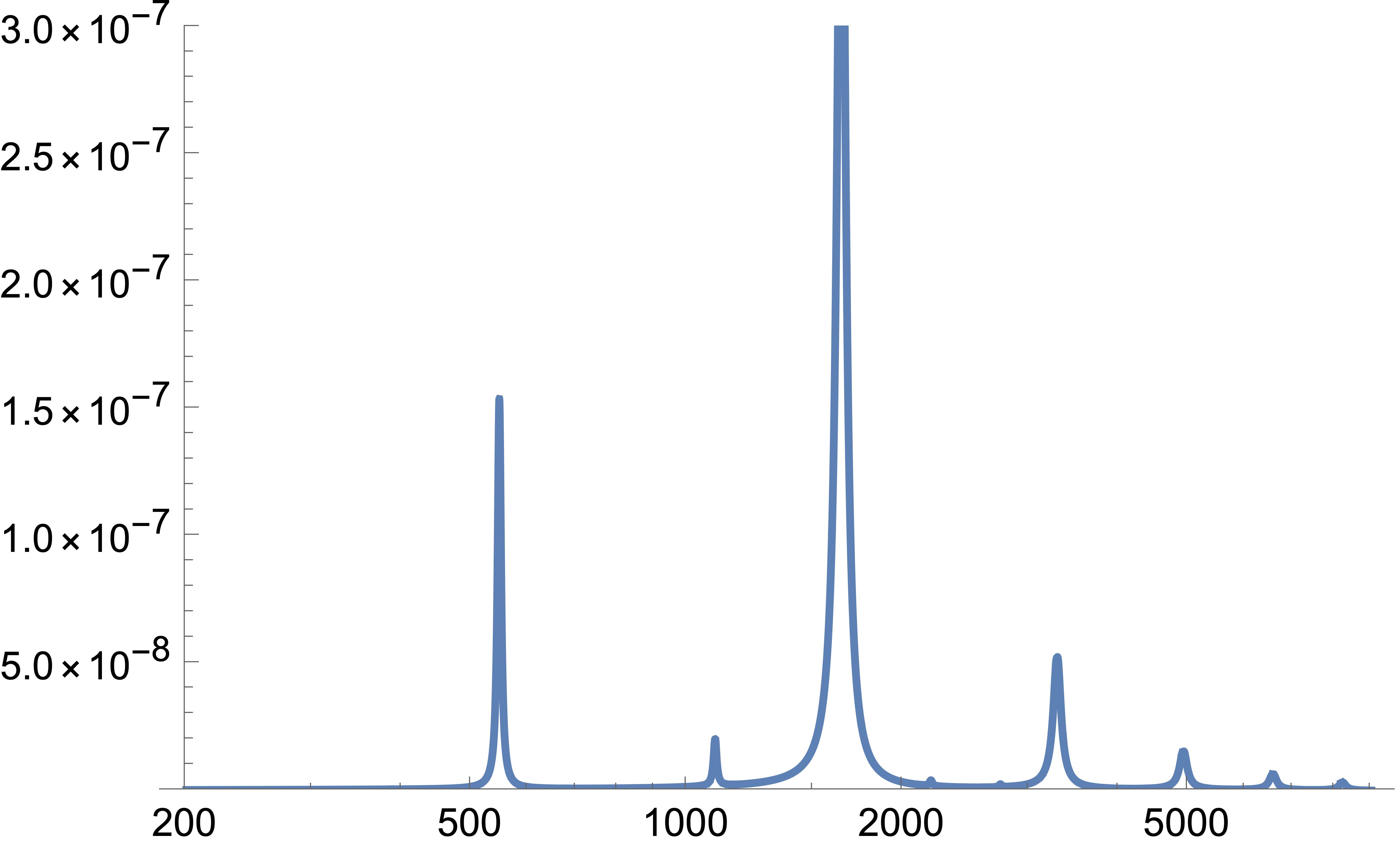}\\[2mm]\includegraphics[width=9truecm]{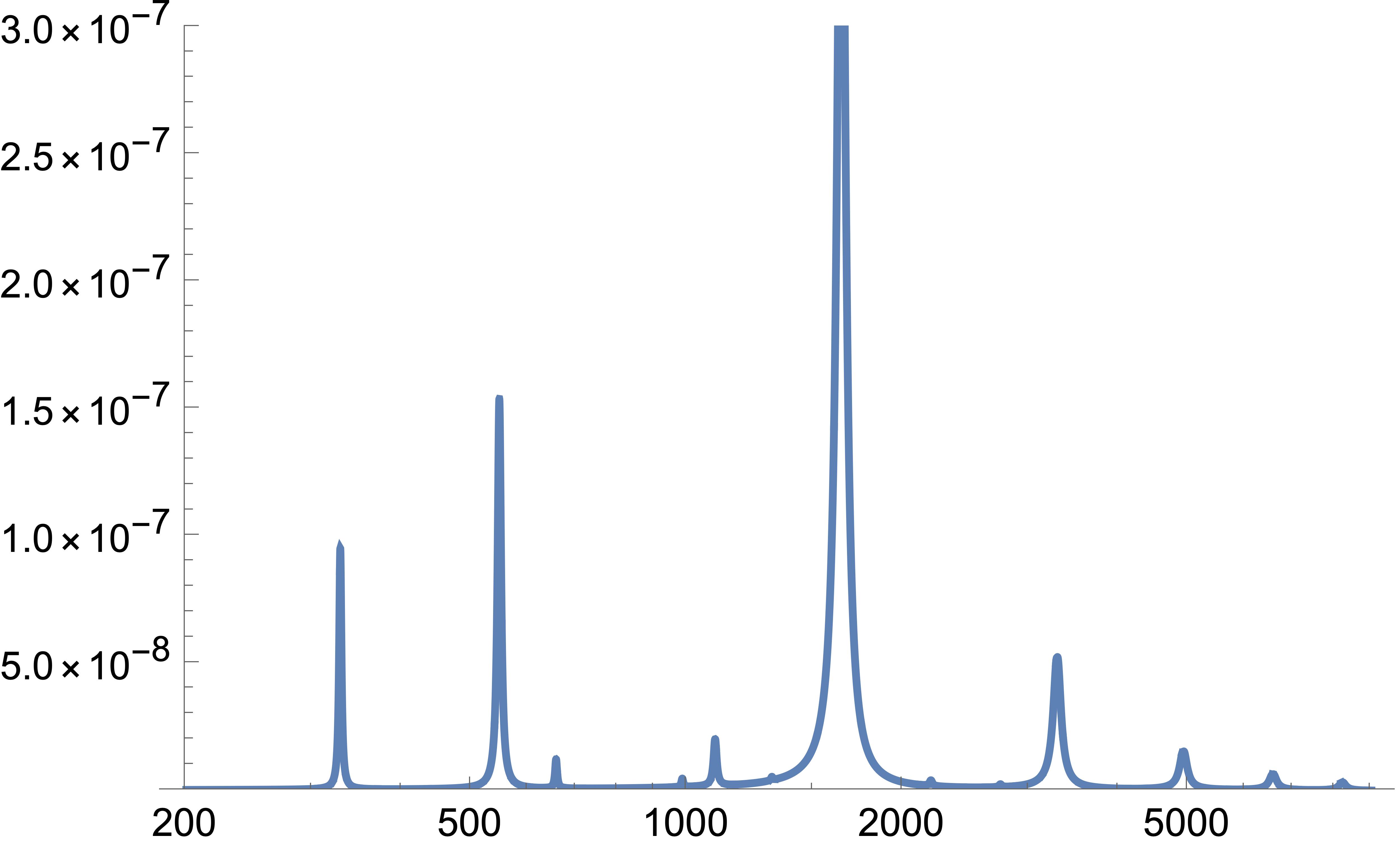}
    \caption{
      $E(\xi,t)$  for a sawtooth corresponding to the C$_4$ for the parameters given in Appendix~\ref{s-parameters}, third column of Table~\ref{t-parameters-original}. For better visibility, in the picture we have fixed $t=0$ (being the variations in $t$ small w.r.t. variations in $\xi$). {\bf First picture:} $E_1(\xi,t)$. Here, one sees the peaks corresponding to the harmonics of the original sound (the highest is the fundamental angular frequency of the sound). {\bf Second picture:} $E_1(\xi,t)+E_2(\xi,t)$. Here, one sees the peaks corresponding to the harmonics of the original sound and of the second sub-harmonics (of the original sound and of its harmonics).
        {\bf Third picture:} $E_1(\xi,t)+E_2(\xi,t)+E_3(\xi,t)$. Here, one sees the peaks corresponding to the harmonics of the original sound and of the second and third sub-harmonics. Notice the presence of sub-harmonics of the different harmonics.
      \label{f-completa}
    }
  \end{center}
\end{figure}

\section{More on the nonlinearity of the response: combination tone\label{s-tartini}}

In this section, we study in more detail $E(\xi,t)$ when the input signal is the superposition of two harmonic sounds with frequencies $\fff_1$ and $\fff_2$ (or, equivalently, with angular frequencies $\kk_1=2\pi \fff_1$ and $\kk_2=2\pi \fff_2$). In other words, we assume that the input signal is the sum of two sinusoidal functions:
\begin{equation}
  \label{due-seni}
  F(t)= c_1 \sin(\kk_1 t+\varphi_1) + c_2 \sin(\kk_2 t +\varphi_2)
\end{equation}
Due to the linearity of the model, such an input produces a response that is the sum of the responses to each individual sinusoidal component and, in particular, only the strings with angluar resonance frequences near $\kk_1$ and $\kk_2$ are significantly activated.
In this section, we show that nevertheless we have the emergence of combination tones. In particular, we show that the energy stored in the strings oscillates with angular frequencies that are linear combinations of $\kk_1$ and $\kk_2$.

Using the linearity of $p_n$ w.r.t. $F(t)$ in \eqref{HOn}, we have that  (cf. \eqref{sol-seno})
\begin{multline*}
  p_n(t)=  c_1 {\cal R}_n(\xi,\kk_1)\sin(\kk_1 t +\varphi_1+\phi_n(\xi,\kk_1)) +\\
  c_2 {\cal R}_n(\xi,\kk_2) \sin(\kk_2 t +\varphi_2+\phi_n(\xi,\kk_2)).
\end{multline*}
Then $E(\xi,t) =\sum_{n=1,2,3,\ldots} E_n(\xi,t) $ where $E_n(\xi,t)$ is given by Formula \eqref{E-general}.
More precisely,
\begin{multline*}
  E_n(\xi,t)=   ( a_1 \sin(\kk_1 t+\alpha_1 ) + a_2 \sin(\kk_2 t+ \alpha_2))^2   \\
  +  ( b_1 \sin(\kk_1 t+\beta_1 ) + b_2 \sin(\kk_2 t+ \beta_2))^2  ,
\end{multline*}
where $a_1,a_2,b_1,b_2,\alpha_1,\alpha_2,\beta_1,\beta_2$ are suitable functions depending on $\xi$ and $n$.
Notice that the terms in $b_1$ and $b_2$ come from the term $\frac{1}{\xi^2 n^2}\dot p_n(t)^2$ in formula \eqref{E-general}. Moreover
the term $\beta_1$ and $\beta_2$ contains a factor $\pi/2$ coming from the fact that $\frac{d}{dt} \sin(k t)=k \cos (kt)=k\sin(kt +\pi/2)$.
Hence, developping the square, we can rewrite $E_n(\xi,t)$ in the form
\begin{equation*}
  E_n(\xi,t)=   f_1(t)+ f_2(t) + f_3(t)  + g_1(t)+ g_2(t) + g_3(t),\nonumber         \\
\end{equation*}
where,
\begin{align}
  f_1(t)= & a_1^2 \sin^2(\kk_1t+\alpha_1),\nonumber                                        \\
  f_2(t)= & a_2^2 \sin^2(\kk_2t+\alpha_2),\nonumber                                        \\
  f_3(t)= & 2 a_1 a_2 \sin(\kk_1t+\alpha_1)\sin(\kk_2t+\alpha_2)\nonumber                  \\
  =       & a_1 a_2\Big[\cos\big((\kk_2-\kk_1)t+\alpha_2-\alpha_1\big)   -\nonumber        \\
          & \qquad\qquad \cos\big((\kk_1+\kk_2)t+\alpha_1+\alpha_2\big)  \Big]   \nonumber
\end{align}
and similarly for $g_1,g_2,g_3$.

We have thus shown that, when the input to the model \eqref{HOn2} is the sum of two sinusoidal functions as in \eqref{due-seni}, $E_n(\xi,t)$ is the sum of  terms oscillating with angular frequencies
\begin{equation}
  2\kk_1,~~2\kk_2,~~\kk_1+\kk_2,~~\kk_2-\kk_1.\label{new-TARTA}
\end{equation}
We stress that this is true even if the model \eqref{HOn2} is linear. The nonlinearity of the response is due to the fact that the signal transmitted to the auditory cortex is not $p_n(t)$ but rather $E_n(\xi,t)$, which is a nonlinear function of $p_n(t)$.

Although a this is outside the scope of this paper, we observe that the presence of these frequencies in the energy stored in the strings is going to produce an effect on the endolymph which in turn produces a feedback to the basilar membrane. This feedback is not described by our linear model, but it is going to  produce additional peaks at the frequencies \eqref{new-TARTA}.
For harmonic sounds with a spectrum composed by several harmonics, the peak around $2\kk_1,$ and $2\kk_2$ are not easily perceived since they are confused with the second harmonics of the two harmonic sounds, while $\kk_2\pm\kk_1$ produces a genuine combination sounds.
Iterating this process, we have the emergence of a whole series of combination tones with frequencies that are linear combinations of the original frequencies $\kk_1$ and $\kk_2$ with integer coefficients.

We stress that this interpretation is coherent with the fact that the combination tone  corresponding to the peak at $\kk_2-\kk_1$ is particularly easy to perceive when $\kk_2=\frac32\kk_1$ (an interval of perfect fifth). Indeed, in this case $\kk_2-\kk_1=2\kk_1-\kk_2$ and hence the combination tone $\kk_2-\kk_1$ coincides with the combination tone between the sound with angular frequency $\kk_2$ and the second harmonic of the sound with angular frequency $\kk_1$, if present.
For the  psychoacoustic literature concerning the combination tone see, for instance,  \citep{Moore2012,caselli2018}.
We also observe that the combination tone $2\kk_1-\kk_2$ emerges also in more complex models, as a byproduct of the nonlinear  amplification of outer hair cells (phenomenon which we are not studying in this paper).  See, for instance,~\cite{goldstein1967auditory,robles1991two,barral2012phantom}.
This amplification mechanism is active primarily at low sound levels. As a consequence, besides the case $\kk_2=\frac32\kk_1$ the combination tone $\kk_2-\kk_1$ is more readily perceived at high sound-pressure levels, whereas the combination tone $2\kk_1-\kk_2$ becomes more salient at low sound-pressure levels

\newcommand{\comb}{\mathrm{combination}}

\section{Concluding remarks}

In this paper, we have shown that the emergence of sub-harmonics and Tartini’s third sound can be explained by a linear model of the basilar membrane, provided that two additional assumptions are made:  {\bf i)}  the auditory cortex receives as input the energy stored in the various strings of the basilar membrane, and {\bf ii)} not only the fundamental mode but also higher oscillation modes must be taken into account.
As a byproduct, phenomena hypothesized as early as the 16th century are shown to be naturally predicted by a simple cochlear model.

Notice that, while nonlinearities are essential to account for phenomena such as Tartini’s third sound, in our model these are incorporated into the quantity transmitted to the auditory cortex and  the underlying mechanical system is represented by a family of linear vibrating strings.
While preserving the necessary non-linear effects, this choice
has the significant advantage of keeping the model explicitly computable, as the mapping from the input signal to the corresponding energy can be written in closed form.

The linear model presented in this work is a first step towards a more complete understanding of the cochlea. Three research directions are particularly interesting. First, it would be interesting to study the feedback model sketched in Section~\ref{s-tartini}, to formally prove the emergence of additional peaks around the frequencies given by \eqref{new-TARTA}, and to compare our predictions with psychophysical data. Second, the model should be extended to account for the interactions between different strings, which are neglected in the present work.
Finally, this linear model should be extended to a fully nonlinear one, which should account for the nonlinear amplification of the outer hair cells. More precisely, the linear wave equation \eqref{string} whose first mode behaves as a damped harmonic oscillator could be replaced by a nonlinear wave equation whose first mode behaves as an oscillator near a Hopf bifurcation, as in \cite{martin2021mechanical}.

\begin{acknowledgments}
  This work has been supported by CNRS through the MITI interdisciplinary programs and partly supported by the ANR-DFG project “CoRoMo” ANR22-CE92-0077-01.
  Ugo Boscain is grateful to Prof. Bernardino Streito from whom he learned  about the problem of the existence of sub-harmonics at the beginning of the 80s and that
  pointed out the references  of Michelangelo Abbado and of Ettore Dabbene.
  The authors are grateful to Pascal Martin, Gilles Monfort and Pietro Polotti for very inspiring discussions.
\end{acknowledgments}

\appendix

\section{Parameters for simulations}
\label{s-parameters}

Under {\bf (A0)}, the model depends on several parameters:   $\ell$, $L$, $A_\rho, A_T,A_\gamma,k_\rho,k_T,k_\gamma$. Although we are mainly interested in a qualitative analysis we try to use values as close as possible to reality.
Unfortunately, to our knowledge, these parameters are not known with great precision and they differ sensibly from one source to another.

Our main references are \citep{enciclopedia} and \citep{nobili2003}. Let us start with $L$ and $\ell$.
\begin{itemize}
  \item From \citep{enciclopedia} we take $L=35mm$.
        Notice that in most literature (as for instance  in \citep{nobili2003}) $L$ is normalized to 1. We make this normalization in the following as well. Notice that with this choice $k_\rho,k_T,k_\gamma$  become adimensional.
  \item Concerning $\ell$ we refer to \citep{enciclopedia}, where $\ell=0.21mm$ at the base $(x=0)$ and  $\ell=0.36mm$. Since we are working under {\bf (A0)}, we are going to fix $\ell=0.29$mm$=2.9 \times 10^{-4}$m independently of $x$.\\[2mm]
\end{itemize}
\newcommand{\nn}{\nonumber}
We collect two possible choices for the other parameters in Table~\ref{t-parameters-original}. The first one is directly deduced from \citep{nobili2003} and the second one is a modification of the first one to have a more reasonable range of frequencies and to fit {\bf (A03)}.
See the following sections for the details on how these parameters have been deduced and modified.

\begin{table}
  \begin{tabular}{|c|c|c|c|}
    \hline
    Param.     & Unit                                                                             & \cite{nobili2003}   & Modif.~Value        \\
    \hline
    \hline
    $A_\rho$   & $\mbox{Kg/m}^2$                                                                  & $3.4\times 10^{-3}$ & $6.9\times 10^{-3}$ \\
    $k_\rho$   & --                                                                               & $2.9$               & $2.3$               \\
    $A_T$      & $\mbox{ Kg/sec}^2$                                                               & $1.5$               & $0.94$              \\
    $k_T$      & --                                                                               & $ 8.5$              & $11.5$              \\
    $A_\gamma$ & $\mbox{ Kg/(m$^2$ sec)}$                                                         & $8.3$               & $8.3$               \\
    $k_\gamma$ & --                                                                               & $-1.6$              & $0$                 \\
    \hline
    \hline
    $\alpha$   & --                                                                               & $0.78$              & $0.33$              \\
    $B_\mu$    & $\mbox{ sec}^{(\alpha-1)}$                                                       & $0.16$              & $24$                \\
    $B_T$      & $\mbox{${\mbox{N}}\,$sec$^{\frac{2k_T}{k_\rho+k_T}}/{\mbox{m}}$}$                & $1.5\times10^{-8}$  & $2.9\times10^{-9}$  \\
    $B_\nu$    & ${{\mathrm{m}^{5/2}}} {\mathrm{sec}}^{\frac{2k_\rho}{k_\rho+k_T}}/{\mathrm{Kg}}$ & $8.7\times 10^{-3}$ & $44\times 10^{-3}$  \\
    \hline
  \end{tabular}
  \caption{Parameters deduced from \citep{nobili2003}, and their modifications in order to have a better fit with the range of audible frequencies.}
  \label{t-parameters-original}
\end{table}

\subsection{Parameters directly deduced from \citep{nobili2003} \label{s-parameters-original}}

The values of the parameters $A_\rho, A_T, A_\gamma$ and $k_\rho, k_T, k_\gamma$ are obtained from \citep{nobili2003} by comparing their model with ours.

More precisely, let us recall equation \eqref{pasta}:
\begin{equation}
  \ddot p_n=  \frac{T}{\rho } \left(-\frac{\pi^2}{\ell^2}n^2 \right)   p_n -\frac{\gamma}{\rho} \dot p_n+\frac{c}{\rho} f_n(t) ,~~~n=1,2,3,\ldots.
\end{equation}
We remark that here  $\rho$ is the density of mass by unit of area (unit of $x$-length and unit of $z$-length). As a consequence, the equation is written in units of mass/length$^2\times$ acceleration namely in  Kg/(m sec$^2$)=N/m$^2$.
Taking the first mode $n=1$ and multiplying it  by $\ell\rho$, we obtain
\begin{equation}
  \ell\rho\, \ddot p_1=  {T}\left(-\frac{\pi^2}{\ell} \right)   p_1 -\ell{\gamma}\, \dot p_1+c\, \ell f_n(t).
  \label{pasta2}
\end{equation}
This can be directly compared with the equation considered in \citep{nobili2003} (see equation {\bf (A1)} in \citep{nobili2003}):
\begin{align}
  m(x) \ddot \xi(x,t) & = -k(x)\xi(x,t)      - [h(x)-\partial_x s(x) \partial_x]\dot \xi(x,t)\nonumber \\
                      &
  +\mbox{external force by unit of length}
  \label{nobiliEQ}
\end{align}
Notice that in equation \eqref{pasta2} $[\ell \rho]$=Kg/m which is the same as $[m(x)]$, and that in \citep{nobili2003} $L$ is normalized to 1, normalization that we make in \eqref{pasta2} as well.

Let us fix $\ell=0.29$mm$= 2.9\times 10^{-4}$m. Comparing \eqref{pasta2}and \eqref{nobiliEQ} we get
$$
  \rho(x)=m(x)/\ell,~~~ T(x)=\frac{\ell}{\pi^2} k(x),~~~\gamma(x)=\frac{h(x)}{\ell}
$$
Since in our model we do not have interaction among the different strings,  we are assuming $s(x)=0$ in equation {\bf (A1)} in  \citep{nobili2003}.

Now
\begin{itemize}
  \item
        In \citep{nobili2003}, Fig.2. $m(x)$ varies between $0.1\times 10^{-5}$ Kg/m (base) to $1.8\times  10^{-5}$ Kg/m (apex). Hence $\rho$ varies between
        $3.4\times 10^{-3}$ Kg/m$^2 $ to $62\times 10^{-3}$ Kg/m$^2$. If we write
        $\rho(x)=A_\rho e^{k_\rho x}$ (with $x=0$ at the base and $x=1$ at the apex) we get
        $$
          A_\rho=3.4\times 10^{-3} \mbox{Kg/m}^2,~~~~k_\rho= 2.9
        $$

  \item
        In \citep{nobili2003}, Fig.2. $k(x)$ varies between $5\times  10^{4}$Kg/(m sec$^2$) (base) to $10^1$ Kg/(m sec$^2$)   (apex).
        Hence $T$ varies between  $1.5$ Kg/sec$^2$ (base) to  $2.9 \times 10^{-4}$  Kg/sec$^2$. Notice that Kg/sec$^2=$N/m.

        If we write
        $T(x)=A_T e^{-k_T x}$ (with $x=0$ at the base and $x=1$ at the apex) we get
        $$
          A_T=1.5\mbox{ Kg/sec$^2$} ,~~~~k_T=8.5
        $$

  \item In \citep{nobili2003}, Fig.2. $h(x)$ varies between $2.4\times  10^{-3}$  Kg/(m sec ) (base) to $0.5 \times 10^{-3}$ Kg/(m sec ) (apex).
        Hence $\gamma$ varies between   $8.3$   Kg/(m$^2$ sec)  (base) to    $1.7$   Kg/(m$^2$ sec)  (apex).

        If we write
        $\gamma(x)=A_\gamma e^{k_\gamma x}$ (with $x=0$ at the base and $x=1$ at the apex) we get
        $$
          A_\gamma = 8.3   \mbox{ Kg/(m$^2$ sec)}  ,~~~~k_\gamma=-1.6.
        $$

\end{itemize}

As a consequence
\begin{align}
  \alpha & :={\frac{2(k_\rho-k_\gamma)}{k_\rho+k_T}}=0.78,\nonumber                                                                                                 \\
  B_\mu  & \equalexpl{under (A0)}\frac{A_\gamma}{A_\rho}   \left( \sqrt{\frac{A_\rho}{A_T}} \frac{\ell }{\pi }    \right)^{\frac{2(k_\rho-k_\gamma)}{k_\rho+k_T}}=\
  0.16  ~ \mbox{sec$^{(\alpha-1)}$},\nonumber                                                                                                                       \\
  B_T    & \equalexpl{under (A0)}A_T   \left( \sqrt{\frac{A_\rho}{A_T}} \frac{\ell }{\pi }  \right)^{\frac{2k_T}{k_\rho+k_T}} =
  1.5\times10^{-8}      \mbox{$\frac{\mbox{N}}{\mbox{m}}$sec$^{\frac{2k_T}{k_\rho+k_T}}$},
  \nn                                                                                                                                                               \\
  B_\nu  & =
  \frac1{A_\rho}   \left( \sqrt{\frac{A_\rho}{A_T}} \frac{\ell }{\pi }    \right)^{\frac{2k_\rho}{k_\rho+k_T}}     \frac{2\sqrt{2}}{\pi}  \sqrt{\ell} =
  8.7\times 10^{-3} \frac{{\mathrm{m}^\frac52}}{\mathrm{Kg}} {\mathrm{sec}}^{\frac{2k_\rho}{k_\rho+k_T}}.\nn
\end{align}

\begin{remark}
  Notice that

  \begin{itemize}

    \item  $[B_\mu]$ are obtained considering that from \eqref{autovalori}

          $[\mu]=[\xi]=$1/sec. Hence from $\mu=B_\mu\xi^\alpha$ we get $[B_\mu ]=$sec$^{(\alpha-1)}$.

    \item  $[B_T]$ are obtained considering that $T(\xi)=B_T\xi^{\frac{2k_T}{k_\rho+k_T}}$ and that $[T]=$N/m and $[\xi]=$1/sec which implies that $[B_T]=\frac{\mbox{N}}{\mbox{m}}$(sec)$^{\frac{2k_T}{k_\rho+k_T}}$

    \item $[B_\nu]$ are obtained considering that  $\nu(x)=\frac{c(x)}{\rho(x)}\frac{2\sqrt{2}}{\pi}  \sqrt{\ell(x)}$. Hence
          $[\nu]=\left[ \frac{ \sqrt{\ell}}{\rho} \right]= \frac{{\sqrt{\mathrm m}}}{{\mathrm{Kg/m}^2}}=\frac{{\mathrm{m}^\frac52}}{\mathrm{Kg}}$.
          From $\nu(\xi)=B_\nu \xi^{\frac{2k_\rho}{k_\rho+k_T}}$ we get $[B_\nu]=\frac{[\nu]}{\left[\xi^{\frac{2k_\rho}{k_\rho+k_T}}\right]} =\frac{{\mathrm{m}^\frac52}}{\mathrm{Kg}} {\mathrm{sec}}^{\frac{2k_\rho}{k_\rho+k_T}}$
  \end{itemize}
\end{remark}

\subsection{Modified parameters from   \citep{nobili2003} with hypothesis  {\bf (A03)} \label{SparametersModifA03} }

Although the parameters given above permit with no difficulties to see the emergence of sub-harmonic series, they do not predict the right range of audible frequencies. Referring to Remark \ref{r-range}, with  the values computed above we get
\begin{align}
  \tilde A & =  2.2 \times 10^5 \mbox{sec}^{-1}\nonumber \\
  \tilde k & =5.7\nonumber
\end{align}
The corresponding frequency range is $[\xi(1)/(2\pi),\xi(0)/(2\pi)]=[120,35\times 10^3]$ Hz which  a bit shifted toward high frequencies w.r.t. the classical range $[20,20\times 10^3]$.

In the following, we propose modified values of $A_\rho, k_\rho, A_T, k_T$ which provide a reasonable range of audible frequencies.
For this reason, in the last column of Table~\ref{t-parameters-original} we report the modified values of these parameters for which
\begin{align}
  \tilde A & =  1.3 \times 10^5 \mbox{sec}^{-1}\nonumber \\
  \tilde k & =6.9\nonumber
\end{align}
The corresponding frequency range is $[\xi(1)/(2\pi),\xi(0)/(2\pi)]=[20,20\times 10^3]$ Hz.

These values are obtained as in the previous section, but considering that
\begin{itemize}
  \item $m(x)$ varies between $0.2\times 10^{-5}$ Kg/m (base) to $2\times  10^{-5}$ Kg/m (apex).
  \item $k(x)$ varies between $3.2\times  10^{4}$Kg/(m sec$^2$) (base) to $.32$ Kg/(m sec$^2$)  (apex).
\end{itemize}
Moreover, to fit hypothesis {\bf (A03)}, we consider  $h(x)=2.4\times 10^{-3}$  Kg/(m sec )  (independently of $x$).
We also take $k_\gamma=0$ to fit hypothesis {\bf (A03)}. See Section \ref{s-physics}.

Notice that the presence of higher damping in this case, renders the peaks less pronounced. They can be rendered more pronounced by lowering $A_\gamma$ which directly influences $B_\mu$ through \eqref{Bmu}.
Notice that the chosen value of $A_\gamma$  used here produces a  bifurcation between the underdamped and the overdamped case (i.e. at
$\xi=\mu(\xi)/2=B_\mu \xi^\alpha/2$) around  $18$Hz, which is close to the lower threshold of the audible frequencies.

\begin{table*}[h]
  \centering
  \caption{Summary of notation}
  \label{tab:parameters}
  \begin{tabular}{|l|l|l|}
    \hline
    \textbf{Parameter}        & \textbf{Description}                                                                                                 & \textbf{Reference}                             \\ \hline\hline
    $x \in [0, L]$            & Length parameter along the cochlea, normalized to $L=1$                                                              & Section \ref{s-physics}                        \\ \hline
    $z \in [0, \ell(x)]$      & Length parameter along the string $x$, $\ell(x)$ is the length of the string $x$                                     & Section \ref{s-physics}                        \\ \hline
    $u^x(t, z)$               & Displacement of the string $x$ at position $z$ and time $t$                                                          & Section \ref{s-physics}, Figure \ref{f-arpa}   \\ \hline
    $\rho(x)$                 & Mass of the string $x$ (per unit length), model: $\rho(x) = A_\rho e^{k_\rho x}$                                     & Section \ref{s-physics}                        \\ \hline
    $T(x)$                    & Tension of the string $x$ (per unit length), model: $T(x) = A_T e^{-k_T x}$                                          & Section \ref{s-physics}                        \\ \hline
    $\gamma(x)$               & Damping of the string $x$. Under {\bf (A0)}: $\gamma(x) = A_\gamma e^{k_\gamma x}$                                   & Section \ref{s-physics}                        \\ \hline
    $c(x)$                    & Coupling of the string $x$ with the external signal, fixed. Under {\bf (A0)}: $c(x) = 1$                             & Section \ref{s-physics}                        \\ \hline
    $F(t, z)$                 & External force (sound). Under \textbf{(A1)}: $F(t, z) = F(t)$                                                        & Section \ref{s-physics}                        \\ \hline
    $\xi$                     & Undamped resonance frequency of the strings.                                                                                                                          
                              & Formula \eqref{xi}                                                                                                                                                    \\ \hline
    $p_n(t)$                  & $n$-th Fourier component of $u^x(t, z)$                                                                              & Section \ref{s-matematica}                     \\ \hline
    $\mu(x)$                  & $\mu(x) = {\gamma(x)}/{\rho(x)}$                                                                                     & Formula \eqref{mum}                            \\ \hline
    $\mu(\xi)$                & $\mu(\xi) = \mu(x(\xi))$ (abuse of notation)                                                                         &                                                \\ \hline
    $\nu_n(x)$                & $\begin{cases}
                                     \frac{c(x)}{\rho(x)}\frac{2\sqrt{2}}{\pi} \sqrt{\ell(x)}\frac{1}{n} & \text{if } n \text{ is odd}, \\
                                     0                                                                   & \text{if } n \text{ is even}
                                   \end{cases}$ & Formula \eqref{nonnepossopiu}                                                    \\ \hline
    $\nu_n(\xi)$              & $\nu_n(\xi) = \nu_n(x(\xi))$ (abuse of notation)                                                                     &                                                \\ \hline
    $\alpha$                  & $\alpha = {2(k_\rho - k_\gamma)}/{(k_\rho + k_T)}$                                                                   & Formula \eqref{Bmu}                            \\ \hline
    $B_\mu, B_\nu, B_T$       & Constants                                                                                                            & Formulas \eqref{Bmu}, \eqref{Bnu}, \eqref{TTT} \\ \hline
    $\kk$                     & Angular frequency of the external sinusoidal signal                                                                  & Formula \eqref{poppa}                          \\ \hline
    $\mathcal{R}_n(\xi, \kk)$ & Asymptotic amplitude of oscillations under a sinusoidal signal                                                       & Formula \eqref{RRn}                            \\ \hline
    $\phi_n(\xi, \kk)$        & Asymptotic phase of oscillations under a sinusoidal signal                                                           & Formula \eqref{PHIn}                           \\ \hline
    $n$                       & Index for sub-harmonics                                                                                              & Section \ref{s-periodic}                       \\ \hline
    $j$                       & Index for harmonics                                                                                                  & Section \ref{s-periodic}                       \\ \hline
    $E(\xi, t)$               & Energy stored in the string with undamped resonance frequency $\xi$                                                  & Section \ref{s-energy}                         \\ \hline
    $E_n(\xi, t)$             & Energy stored in the $n$-mode of the string with undamped resonance frequency $\xi$                                  & Section \ref{s-energy}                         \\ \hline
  \end{tabular}
\end{table*}

\bibliography{0-0-references}

\end{document}